# Estimating unrestricted spatial interdependence in panel spatial autoregressive models with latent common factors[1]

Deborah Gefang

Department of Economics, Leicester University

Stephen G Hall

Department of Economics, Leicester University, Bank of Greece,

and University of Pretoria

George S. Tavlas

Bank of Greece and the Hoover Institution, Stanford University

## Abstract

We develop a new Bayesian approach to estimating panel spatial autoregressive models with a known number of latent common factors, where $N$, the number of cross-sectional units, is much larger than $T$, the number of time periods. Without imposing any a priori structures on the spatial linkages between variables, we let the data speak for themselves. Extensive Monte Carlo studies show that our method is super-fast and our estimated spatial weights matrices and common factors strongly resemble their true counterparts. As an illustration, we examine the spatial interdependence of regional gross value added (GVA) growth rates across the European Union (EU). In addition to revealing the clear presence of predominant country-level clusters, our results indicate that only a small portion of the variation in the data is explained by the latent shocks that are uncorrelated with the explanatory variables.

---

[1] An earlier version of this paper appeared as Identifying spatial interdependence in panel data with large N and small T

# 1. Introduction

Since the seminal paper by Cliff and Ord (1973), spatial autoregressive (SAR) models have been widely used in the literature to investigate the cross-sectional interdependencies between variables (e.g., Anselin 1988, Baltagi et al. 2003; 2013, Lee and Yu 2010, LeSage and Pace 2018, to mention a few). In a SAR model, the spatial structure between units is captured by the $N \times N$ spatial weights matrix with zero diagonal entries. Since it is often challenging to estimate the $N^2 - N$ off-diagonal elements in a spatial weights matrix, most of the available studies elicit the matrix a priori with references to theories and conventions.

As summarized in Fingleton and Arbia (2008), a major criticism of SAR models is that their inferences such as the spillover effects are sensitive to how the spatial weights matrix is specified. There are two major drawbacks of imposing the predetermined spatial weights matrices to a model without letting the data speak. To start with, economic theories and conventions are not able to provide researchers with precise spatial weights matrix entries. In addition, for the same data, different theories and conventions may imply very different spatial linkage relationships, which may even contradict each other. Studies such as those by Kelejian (2008) and Kelejian and Piras (2011, 2016) have developed various advanced testing methods to select the 'true spatial weights matrix' among a number of plausible candidates. More recently, Higgins and Martellosio (2023) develop a penalised quasi-maximum likelihood estimator, using adaptive Lasso to purge the irrelevant predetermined weighting matrices. However, the success of the selection procedure depends on the 'true spatial weights matrices', which are hard to pin down, being included in the candidates pool in the first place. Hence, it is important to work with models in which the spatial weights matrices are not set a priori.

Early efforts to estimate the spatial weights matrices usually relied on imposing less restrictive assumptions on the spatial structure (e.g., Pinkse et al. 2002, Bhattacharjee and Jensen-Butler 2013,

Bailey et al. 2016). In recent years, developments in various variable selection and parameter shrinkage techniques suitable for estimating models where the number of parameters is larger than the number of observations have equipped researchers with tools to estimate the spatial weights matrices directly from panel data. Using these developments, Ahrens and Bhattacharjee (2015) introduce a two-step Lasso estimator to uncover an unrestricted spatial weights matrix. Lam and Souza (2020) propose to combine prior knowledge on spatial interdependence with data information captured by an adaptive Lasso estimator of Zou (2006). Gefang et al. (2023) use a two-stage variational Bayesian (VB) approach with Dirichlet-Lasso (D-L) and Horseshoe priors to estimate the unrestricted spatial weights matrix. Krisztin and Piribauer (2023) introduce a Bayesian approach to estimate a spatial weights matrix with its off-diagonal elements being either 0 or 1 before being row-standardised. More recently, in the network literature, de Paula et al. (2023) have derived conditions under which the unrestricted social interactions matrix and its associated social effect parameters are globally identified.[2] They use adaptive elastic net generalized methods of moments of Caner and Zhang (2014) for estimations.

In comparison with the methods proposed in Ahrens and Bhattacharjee (2015) and Gefang et al. (2023), methods of Lam and Souza (2020) and Krisztin and Piribauer (2023) can be less appealing as the former only works when the spatial linkages proposed by the researchers are not too different from the true one, while the latter rules out the possibility of any negative elements in a spatial weights matrix. In addition, both Lam and Souza (2020) and Krisztin and Piribauer (2023) assume the same spatial parameters for all units, while Ahrens and Bhattacharjee (2015) and Gefang et al.

---

[2] SAR models of various forms are widely used to analyse social and economic networks (e.g. Manski 1993, Diebold and Yilmaz 2014, de Paula et al. 2023, and Huang et al. 2023). In these studies, a spatial weights matrix is usually called the social interactions matrix to reflect the fact that the matrix is used to capture the network effect.

(2023) do not impose such restrictions. de Paula et al. (2023) do not impose restrictions on the spatial weights matrix either, but they assume the same spatial parameters for all units.

Following Ahrens and Bhattacharjee (2015) and Gefang et al. (2023), this paper focuses on spatial panel data models with an unrestricted spatial weights matrix in order to let the data speak. More specifically, we develop a simple two-step VB approach to estimating spatial panel data model where the ratio of $N$ to $T$ is much larger than those that can be dealt with in Ahrens and Bhattacharjee (2015) and Gefang et al. (2023).[3] This advancement is nontrivial, as their methods become computationally prohibitive when N>100, whereas our approach can efficiently estimate models with N as large as 500.

Papers such as Hsiao and Pesaran (2008) emphasize the importance of accounting for heterogeneity across units and over time when specifying coefficients and error structures in panel data models. This insight is equally relevant to the specification of panel spatial models. In the spatial econometrics literature, researchers typically employ random effects, fixed effects, or seemingly unrelated regression (SUR) models to capture heterogeneity, as discussed in the survey paper by Elhorst (2003). More recently, Higgins and Martellosio (2023) introduce a factor structure in the error terms to account for unobserved heterogeneity. Their approach is particularly appealing, as the factor structure can identify common shocks affecting different units and thereby provide deeper insights into the data-generating process. To better address heterogeneity arising from various sources, this paper introduces a model that allows for heterogeneous coefficients across individual units, and incorporates a small known number of unobserved latent factors in the error terms in the fashion of Higgins and Martellosio (2023).

---

[3] Various variable-selection and parameter-shrinkage techniques can be employed to estimate the unrestricted parameters when the number of parameters exceeds the number of observations. Excellent review papers include Korobilis and Shimizu (2022) for Bayesian methods and Yoo (2024) for frequentist approaches.

We use D-L priors for variable selection and parameter shrinkage. Despite the caveat that global-local-shrinkage priors tend to shrink all the non-diagonal spatial weights entries towards zero, but never to zero, when $N$ is much larger than $T$, our estimated spatial weights matrices turn out to be highly similar to their true counterparts. This feature is of particular importance to researchers, because, as shown in Lesage and Pace (2014), drastically different choices of spatial weight matrices can lead to substantially different estimation and inference results, whereas when two spatial weight matrices are similar, the estimated spatial effects are relatively insensitive to the choice of matrix.

Extensive Monte Carlo exercises show that our Bayesian approach works well. Keeping $T$ as small as 20, we are able to uncover the spatial relationships for $N$ as large as 500 (which implies 249,500 spatial parameters are to be estimated), while simultaneously identifying the true latent factors in the error terms.

Noting the limitation that traditional spatial weights matrices only contain non-negative entries, we specifically allow for both positive and negative spatial weights in the simulations. Monte Carlo results show that our estimated spatial weights matrices are similar to the true ones. Interestingly, keeping $T$ unchanged, while the simple correlations[4] between the estimated and true spatial

---

[4] Simple correlation between two matrices $A$ and $B$ is

$$corr2 = \frac{\sum_n \sum_m (A_{nm} - \bar{A})(B_{nm} - \bar{B})}{\sqrt{(\sum_n \sum_m (A_{nm} - \bar{A})^2)(\sum_n \sum_m (A_{nm} - \bar{A})^2)}}.$$

weights matrices decreases while $N$ increases, the structural similarity index measure (SSIM)[5] between those two matrices goes up, indicating when noise is removed, large $N$ is actually helpful in uncovering the true spatial weights matrices.

We also compare and contrast between the true direct effects and indirect effects, which are used to measure the spillovers between variables. As we briefly mentioned above, our Monte Carlo results are very encouraging in the sense that the estimated direct effects are almost identical to the true direct effects, and the estimated indirect effects are highly correlated with the true indirect effects. Despite the drawbacks that, when $N$ becomes larger, the magnitudes of the estimated indirect effects tend to be much smaller than their true values, at the minimum, we can use the estimated effects matrices to identify the spillover patterns between variables with confidence.

For the empirical application we examine the spatial interconnections and spillovers between European Union (EU) regional gross value added (GVA) growths. Due to its important economic and policy implications, spatial interdependence of regional economies has been intensively investigated since early 2000s. A majority of the studies set a priori spatial weights matrices based on geographical distances and social economic proximities (e.g., Arbia et al. 2010, Basile et al. 2014, Crespo et al. 2014, and Piribauer 2016, to mention a few). Estimating spatial weights matrix from the data, an interesting study by Piribauer et al. (2023) finds significant country clusters and marked divide between regions in Western and Northern Europe and those in Southern and Eastern Europe. The spatial weights matrix used in that study, however, can nevertheless be considered as quite restrictive. To start with, the authors assume that all the non-zero entries in the same row of the spatial weights matrix are of the same value, implying all regions spatially related to an

---

individual region shall have the same spillover effects on that region, which can be unrealistic. Furthermore, the entries of their spatial weights matrix are all non-negative, making it impossible to account for phenomenon in which a region's developments hollow out its spatially related neighbours.

Using an unrestricted spatial weights matrix, we find pronounced country clusters dominating other spatial connections, as in Piribauer et al. (2023). But the country groupings we uncover are more complicated than previously known. To start with, we find that a small number of regions are associated with negative spatial weights, implying their economic growth might be attracting resources from other regions, at the latter's expense. In addition, we find that, in comparison with other major EU economies, the German economy is much less linked with other country's economies. Moreover, instead of the well documented divides, such as the north-south and west-east divides, between country groups, we find regions in countries such as France, Spain, Portugal and Italy share clear spatial linkages. We also find a number of other interesting spatial interdependencies which have potential important policy implications. For example, the Greek economy is closely spatially affected by Spain's, but much less so by other countries.

The rest of the paper is organised as follows. Section 2 describes the unrestricted panel SAR model with latent common factors and the Bayesian estimation approach. Section 3 conducts Monte Carlo exercises. Section 4 uses an application on EU GVA data to show how our modeling framework and estimation technique can shed important new light on a topic that has been extensively investigated in the literature. Section 5 concludes.

## 2. **The Model and Bayesian Techniques**

### 2.1 SAR with Factor Structure

The spatial panel data model is assumed to take the following simple form:

$$
\begin{bmatrix} y_{1t} \\ y_{2t} \\ \vdots \\ y_{Nt} \end{bmatrix} = \begin{bmatrix} 0 & W_{12} & \dots & W_{1N} \\ W_{21} & 0 & \dots & W_{2N} \\ \vdots & \vdots & \dots & \vdots \\ W_{N1} & W_{N2} & \dots & 0 \end{bmatrix} \begin{bmatrix} y_{1t} \\ y_{2t} \\ \vdots \\ y_{Nt} \end{bmatrix} + \begin{bmatrix} x_{1t,1} & \dots & x_{1t,k} & 0 & \dots & 0 & \dots & 0 & \dots & 0 \\ 0 & \dots & 0 & x_{2t,1} & \dots & x_{2t,k} & \dots & 0 & \dots & 0 \\ \vdots & \dots & \vdots & \vdots & \dots & \vdots & \dots & \vdots & \dots & \vdots \\ 0 & \dots & 0 & 0 & \dots & 0 & \dots & x_{Nt,1} & \dots & x_{Nt,k} \end{bmatrix} \begin{bmatrix} \theta_{1,1} \\ \vdots \\ \theta_{1,k} \\ \theta_{2,1} \\ \vdots \\ \theta_{2,k} \\ \vdots \\ \theta_{N,1} \\ \vdots \\ \theta_{N,k} \end{bmatrix} + \begin{bmatrix} e_{1t} \\ e_{2t} \\ \vdots \\ e_{Nt} \end{bmatrix} \quad (1)
$$

$$
y_t = W y_t + x_t \theta + e_t
$$

where $y_t$ is an $N \times 1$ vector of dependent variables, and $x_t$ is the $N \times Nk$ matrix of exogenous explanatory variables (and possibly some lagged variables), which may include an intercept and dummy variables capturing time-specific effects. The $N \times N$ spatial weights matrix $W$ (with zero diagonal entries), and the coefficients $\theta$ are to be estimated from the data.

The sum of each row in the unstandardised $W$ can be treated as a spatial parameter that is associated with the individual cross sectional unit if we use the row-standardised W as the spatial weights matrix, as is common in the literature. In this setting, each unit has its own spatial parameter, analogous to random coefficients that capture individual-specific effects, as discussed in Hsiao and Pesaran (2008). However, unlike Hsiao and Pesaran (2008), we do not impose any particular structure on these spatial parameters beyond the requirement of invertibility, making our model more flexible.

In reality, there might be more than one type of spatial linkages between the cross sectional units of $y_t$. In that case, Model (1) can be viewed as a reduced form model, where $W$ is the linear or nonlinear combination of individual spatial weights matrices. Spatial linkages may also exist between exogenous explanatory variables of different units. In such cases, instead of adopting a block-diagonal structure, $x_t$ should be arranged in a denser form to allow $x_{jt}$ to have impact on $y_{it}$, whenever spatial linkages are present between $x_{it}$ and $x_{jt}$.

To account for unobserved heterogeneity that can be attributed to a kown number common shocks, we further assume the error term $e_t$ to take the following factor structure form:

$$e_t = \Lambda f_t + \varepsilon_t \qquad (2)$$

where $f_t$ is a $l$ by 1 vector of common factors, and $\Lambda$ is a $n$ by $l$ factor loading matrix. For identification, we assume $f_t$ and $x_t$ are uncorrelated. The distribution of the error terms are assumed to be $\varepsilon_{it} \overset{iid}{\sim} N(0, \sigma_i^2)$. [6]

## 2.2 Bayesian Estimation Techniques

If the researchers have more knowledge in the true spatial weights matrices, they can make use of that knowledge in various ways. For example, one can set some tight priors to allow for small deviations from the predetermined spatial weights matrices like Lam and Souza (2020), or using shrinkage methods to select the more likely spatial weights matrices from a large number of candidates following Higgins and Martellosio (2023). These restrictions, based on sensible insights, will enhance the model's performance. However, it is not always easy to come up with all the true weights matrices. The focus of the current paper, in line with Ahrens and Bhattacharjee (2015) and Gefang et al. (2023), is to uncover the spatial weights matrix providing that we have no prior information of the true values of its elements, particularly in the presence of additional model complexities such as factor structures in the error terms.

Given that $f_t$ and $x_t$ are uncorrelated, frequentist studies such as Chang et al. (2015) and Gao and Tsay (2024) have demonstrated that the estimation of W and θ can be treated as a standard least squares problem. Building on their insights, we adopt a two-phase estimation procedure: In the first phase, W and θ are estimated using a simple 2-step VB estimation approach; In the second phase, we identify the latent factors $f_t$ and loadings Λ from the residuals obtained in the first phase.

---

[6] The error terms can follow distributions other than the Normal. For example, one can use t-distribution to allow for heavier tails. In such cases, the priors and posteriors of $\sigma_i^2$ introduced in Section 2.2 would need to be adjusted accordingly.

### 2.2.1 Estimating W and θ using 2-step VB

As Piribauer et al. (2023) rightly observed, despite the fact that the methods of Ahrens and Bhattacharjee (2015) and/or Gefang et al. (2023) are able to estimate the large number of free parameters involved in model (1), their approaches are not suitable for panel data where $N \gg 200$ while $T$ remains rather small. Since panel data are often featured by large $N$ and small $T$, it is important to develop an approach to estimating model (1) that can surmount the above mentioned limitations. Keeping that in mind, in what follows we propose a simple two-step VB estimation approach to estimating W and θ in equation (1).

First, we use all the exogenous (or predetermined) variables contained in $x_t$, namely $x_{1t,1}, \dots, x_{1t,k}, x_{2t,1}, \dots, x_{2t,k}, \dots, x_{Nt,1}, \dots x_{Nt,k}$, as instrumental variables to calculate the predicted value of $y_{it}$, for $i = 1, \dots, N$, one at a time. Second, estimate model (1) equation by equation: For the $i^{th}$ equation, use $y_{it}$ as the dependent variable but replace the remaining endogenous variables $y_{jt}$s (for $j = 1, \dots, N$ and $j \neq i$) on the right-hand side of the equation by their predicted values derived in the first step. In essence, those two steps are similar to the two-stage least squares (2LSL) in which the parameters can be estimated equation by equation. This approach is attractive because the estimations can be performed in parallel for each step.

The method is similar to that of Gefang et al. (2023). The only difference is that in the first step, Gefang et al. (2023) estimated the predicted endogenous variables using a big vector autoregressive model (VAR), assuming the variance covariance matrix of the VAR error terms to be non-diagonal. Their VAR framework correctly accounts for the interrelationship between endogenous variables, yet it slows down the computations, hence making estimating models of larger number of cross-sectional units, such as those with $N \gg 200$, infeasible. In this paper, we instead estimate the predicted endogenous variables equation by equation. Although this approach sacrifices the nice VAR feature, it gains us huge advantage in computing speed.

We use D-L prior of Bhattacharya et al. (2015) for variable selection and parameter shrinkage. As with other popular global–local shrinkage priors (e.g., Polson and Scott 2012a; 2012b, Carvalho et al. 2010), the computing cost of D-L priors is relatively low, hence suitable for estimating models with a large number of parameters. Moreover, with D-L priors, the entire posterior distribution concentrates at the optimal rate (Bhattacharya et al., 2015). Finally, studies such as Zhang and Bondell (2018) have shown that D-L prior leads to posterior consistency and selection consistency.[7]

In both steps, each single equation to be estimated can be written in the following general form:

$$y = X\beta + e \tag{3}$$

where $y$ is a $T \times 1$ vector of dependent variables, $X$ is a $T \times M$ matrix of explanatory variables, and $e$ is a $T \times 1$ vector of *iid* error terms with $e_t \sim N(0, \sigma^2)$.

The hierarchical D-L priors for the $m^{th}$ element, for $m \in (1, ..., M)$, of $\beta$ are as follows:

$$\begin{aligned} \beta_m \mid \phi_m, \tau &\sim DE(\phi_m \tau) \\ \phi_m &\sim Dir(a, ..., a) \end{aligned} \tag{4}$$

where $DE$ denotes the Double Exponential or Laplace distribution, and $Dir$ is the Dirichlet distribution. Finally, we set the following Gamma priors for $\tau$ and $\sigma^{-2}$:

---

[7] Nevertheless, other global–local shrinkage priors can be used instead of D-L for the simple two-stage VB, depending on the researchers' preferences.

$$\tau \sim G(Ma, \tfrac{1}{2})$$
$$\sigma^{-2} \sim G(\underline{v}, \underline{s}) \tag{5}$$

The above hierarchical D-L priors for $\beta$ can be expressed as

$$\beta \sim N \left( \begin{bmatrix} 0 \\ 0 \\ \vdots \\ 0 \end{bmatrix}, \begin{bmatrix} \varphi_1 \phi_1^2 \tau^2 & 0 & \ldots & 0 \\ 0 & \varphi_2 \phi_2^2 \tau^2 & \ldots & 0 \\ \vdots & \vdots & \ddots & \vdots \\ 0 & 0 & \ldots & \varphi_M \phi_M^2 \tau^2 \end{bmatrix} \right) \tag{6}$$

and

$$\varphi_m \sim Exp(\tfrac{1}{2}) \tag{7}$$

Note that for the D-L priors, the only prior we need to select is $a$. This makes the prior sensitivity analyses a lot easier. With Normal prior for $\beta$ and Gamma prior for $\sigma^{-2}$, their posteriors can be drawn by Gibbs sampling.

When the number of parameters is large, as in our case, Gibbs sampling can become computationally expensive or even infeasible. To address this issue, we employ VB as a computationally efficient alternative. As described by Blei et al. (2017), VB approximates the posterior with a simpler density that is as close as possible to the true posterior in the Kullback–Leibler sense. Studies such as Gefang et al. (2019, 2020) have shown that VB performs well for large datasets, particularly in models using global–local shrinkage priors. For brevity, we present the conditional posteriors and VB densities in the Online Appendix. Incorporating VB, we adopt the following simple two-step algorithm to estimate the parameters in model (1):

---

**Simple two-step VB Algorithm**

---

**1. Calculate the predicted values of the endogenous variables**

For $i = 1, ..., N$, run in parallel to predict $y_{it}$ using all the exogenous variables as instrumental variables

- Initialize the parameters and hyperparameters
- Run VB until the changes in parameters are negligible
- Save the predicted values of $y_{it}$

End

**2. Estimate model (1) equation by equation**

For $i = 1, ..., N$, substitute the endogenous variables on the right hand side of the equation with the predicted values derived in the first step, run in parallel to estimate the $i^{th}$ row of $W$ and $(\theta_{i1}, ... \theta_{ik})'$.

- Initialize the parameters and hyperparameters
- Run VB until the changes in parameters are negligible
- Save the parameters

End

---

## 2.2.2 Estimating Latent Factors and Loadings

Substituting the estimated W and θ into equation (1), we end up with the following factor model:

$$\hat{e}_t = f_t \Lambda + \varepsilon_t \tag{8}$$

where $\hat{e}_t = (I - W)y_t - x_t\theta$, $\varepsilon_t \sim N(0, \Sigma)$, and $\Sigma = diag(\sigma_1^2, ..., \sigma_n^2)$

Following Chan et al. (2018), we write equation (8) in matrix form as $\hat{e} = F\Lambda + \varepsilon$, where $\hat{e}$ and $\varepsilon$ are $T$ by $N$ matrices with $E(vec(\varepsilon)vec(\varepsilon)') = \Sigma \otimes I_T$, $F$ is the $T$ by $l$ matrix such that the product $F\Lambda$ is of rank $l$.

Assigning uniform priors to the components of the singular value decomposition (SVD) of $F\Lambda$, Chan et al. (2018) introduced the following ordering invariant prior distribution for $F$ and $\Lambda$:

$$p(F,\Lambda) \propto \exp\{-\frac{1}{2}tr(F'F) - \frac{c_\lambda}{2}tr(\Lambda'F'F\Lambda)\} \tag{9}$$

where $tr(A)$ indicating the trace of the square matrix $A$, and $c_\lambda$ governs the degree of shrinkages towards zero. Multiplying the likelihood function and the priors, we are able to derive the conditional posteriors for $F$ and $\Lambda$:

$$vec(F)|vec(\Lambda) \sim N(0, (I_m + c_\lambda \Lambda\Lambda^{-1})^{-1} \otimes I_T) \tag{10}$$

$$vec(\Lambda)|vec(F) \sim N(0, \frac{I_k}{c_\lambda} \otimes (F'F)^{-1}) \tag{11}$$

Setting inverse Gamma priors for $\sigma_i^2$, its conditional posterior is also inverse Gamma.

As noted in Chan et al. (2018), model comparison methods such as Savage-Dickey density ratio can be used to select $l$, the number of latent variables in the model (i.e., the rank of $F\Lambda$). In this paper, however, we assume that $l$ is known based on the model setup or theoretical considerations, in order to simplify the computations.

## 3. Monte Carlo Experiments

Extensive Monte Carlo studies have shown that the simple two-step VB works well for spatial panel models where $N \gg T$. To give some flavor, we report the estimated results of three simulated datasets where $T = 20$, and $N = 30$, $300$ and $500$ respectively. To evaluate the performance of two-step VB algorithm under models of varying complexity, for fixed $N$ and $T$, we consider spatial models both with and without factor structures in their error terms.

In all cases, we construct $W$, the true spatial weights matrix, in three steps. First, we use the '$q$ ahead and $q$ behind' method of Kelejian and Prucha (1999) to establish the basic relationship matrix to specify that each unit is only spatially associated with the $q$ units in front of it and the $q$ units after it. Specifically, for N = 30, 300, and 500, we set q =13, 140, and 230, respectively. We then update the non-zero elements in the basic relationship matrix by values randomly drawn from the Normal distribution $N(0,1)$, and standardise the basic relationship matrix so that the sum of each of its rows is equal to 1. Second, we generate the spatial parameter $\rho_i$ associated with cross-sectional unit $i$ by a random number uniformly drawn from the open interval (0, 1). Finally, we multiply each row of the basic spatial relationship matrix by its unit specific spatial parameter to get $W$. The elements of the coefficients $\theta$ are drawn from the Uniform distribution U(0,1) and the elements of factor loadings $\Lambda$ are drawn from $N(0,1)$.

Holding the true parameters $W, \theta$ and $\Lambda$ fixed, we generate the exogenous variables $x_{it}$, the latent factors and the *iid* disturbances from $N(0,1)$ over 1,000 Monte Carlo replications.

Our two-step VB algorithm is super fast. On a laptop[8], each stage takes around 2 seconds to estimate for panel data models where $N = 30$, around 2 minutes to estimate for models where $N = 300$, and around 10 minutes to estimate for models where $N = 500$.

### 3.1 The Spatial Weights Matrix

---

We present the heatmaps of the true spatial weights matrices on the left-hand side of the panels in Figures 1–6, and the corresponding mean estimated spatial weights matrices on the right-hand side.[9] By visual inspection, the corresponding heatmaps on the left and right exhibit similar spatial patterns, with the estimated matrices more closely resembling the true ones when the model does not include latent factor structures in the error terms. A closer look, however, reveals that as N increases, the magnitudes of the estimated values corresponding to the true nonzero elements tend to shrink globally toward zero, which is not ideal. Nevertheless, although the estimated values for the true zero elements are seldom exactly zero, they are generally much closer to zero than those associated with the true nonzero elements.

A natural question that arises is to what degree the estimated spatial weights matrices resemble the true spatial weights matrix. To measure the similarities between the estimated spatial weights matrices and their corresponding true weights, we report their simple correlation and SSIM of Zhou et al. (2004) in Tables 1–2.

Table 1. Similarities between the Estimated and True Spatial Weights Matrices, without Factors

|  | corr2 | SSIM |
| --- | --- | --- |
| N=30 | 0.9966 | 0.8904 |
| N=300 | 0.9844 | 0.9877 |
| N=500 | 0.9772 | 0.9957 |

Table 2. Similarities between the Estimated and True Spatial Weights Matrices, with Factors

---

[9] For $N = 300$ and $N = 500$, we only plot their first $30 \times 30$ submatrices in order to make the figures legible.

|       | corr2  | SSIM   |
|-------|--------|--------|
| N=30  | 0.7770 | 0.8182 |
| N=300 | 0.5572 | 0.9876 |
| N=500 | 0.8114 | 0.9957 |

Results of simple correlation show that the estimated spatial weights matrices more closely resemble their true counterparts in models without factor structures (the less complex specification) than in models with factor structures (the more complex framework). However, SSIM measures indicate that adding factor structure in the model doesn't have as big of an impact on the similarities between the estimated and true weighting matrices, especially when $N$ gets larger.

Both simple correlation and SSIM are positive and in the range from .5572 to .9966 (correlation) and from .8182 to .9957 (SSIM). Consistent with the messages revealed in Figures 1–6, the value of simple correlation between the estimated spatial weights matrix and the true spatial weights matrix tends to decrease as $N$ increases. However, the structural similarities between the estimated spatial weights matrix and its corresponding true spatial weights matrix measured by SSIM increases when $N$ increases. Despite simple correlation results indicate that there are less similarities between the estimated matrix and the true one when N gets larger, SSIM show that taking out the impacts of noises, when N gets bigger, the estimated spatial weights matrix is actually becoming more similar to its true counterpart. This finding is particularly encouraging as it implies that we can rely on the estimated spatial weights matrix to make more accurate inferences.

### 3.2 Effects Estimates

LeSage and Pace (2018) argue that effects estimates shall be the focus of Monte Carlo exercises because applied practitioners are more interested in the direct and indirect effects of changes in an explanatory variable on the dependent variable than any individual parameters. In the spirit of LeSage and Pace (2009), we define the effects matrix of model (1) as

$$\partial y / \partial X_r = (I_N - W)^{-1} diag(\theta_{ir}, ..., \theta_{Nr}), \qquad (12)$$

where $r \in [1, ..., k]$. The diagonal elements of the partial derivative matrix, which reflects the own partial derivative, $\partial y_i / \partial X_{i,r}$, measure the direct effect, and the off-diagonal elements of $\partial y_i / \partial X_{i,r}$, by contrast, measure the corresponding indirect effects.[10]

In Monte Carlo, we calculate $\partial y / \partial X_r$ for $r = 1, 2$. The findings from $\partial y / \partial X_1$ and $\partial y / \partial X_2$ are qualitatively the same. To save space, we focus on examining the former, and henceforth call it the effects matrix, which measures the spillover effects of a one unit increase in $X_1$.

Since the direct effects are much larger than the indirect effects, we display the diagonal elements of the true effects matrices in the top-left panels of Figures 7–12, and the off-diagonal elements in the corresponding top-right panels. For ease of comparison, the diagonal elements of the estimated effects matrices are shown in the bottom-left panels, while the off-diagonal elements are shown in the bottom-right panels.

Two important messages emerge from Figures 7–12. First, regardless whether $N$ is large or small, the estimated direct effects remain very similar to the true direct effects. Second, the larger is $N$, the smaller is the magnitude of the indirect effects in comparison with the true indirect effects, however, the estimated indirect effects matrix and its true counterparts seem to share the same amount of similarities regardless of the value of $N$, or whether the factor structure is present in the error terms or not. The first finding is not surprising as

$(I_N - W)^{-1} = I_N + W + (W)^2 + \cdots$. With the infinity norm of $W$ being less than 1, $I_N$ plays a

---

[10] In LeSage and Pace (2009), the direct effects are defined as the average of the diagonal elements and the indirect effects are defined as the average of the off-diagonal elements. By contrast, we focus on each individual element of the partial derivative matrix.

dominant role in calculating the direct effects, and the impact of $W$ only takes place via the diagonal elements of $(W)^2$, $(W)^3$ and so on, which decrease rapidly. The second finding, however, is more interesting. To further investigate what happens, we report the similarities between the true and estimated effects matrices in Tables 3–4. Indeed, the estimated and true direct effects matrices are almost equivalent in all cases with both simple correlation measures and SSIM approaching 1. Measured by simple correlation, the estimated and true indirect effects matrices are highly correlated, especially for models without latent factors. It is striking to observe that for models with factor structures, corr2 measures for indirect effects only drop slightly when $N$ increases from 30 to 300, then goes up when $N$ is 500. SSIM measures for indirect effects tend to go up when $N$ increases, similar to SSIMs reported for the true and estimated spatial weights matrices.

Table 3. Similarities between the Estimated and True indirect Effects Matrices, without Factors

|       | Direct Effects | | Indirect Effects | |
| --- | --- | --- | --- | --- |
|       | corr2 | SSIM | corr2 | SSIM |
| N=30  | 1.0000 | 1.0000 | 0.9978 | 0.9374 |
| N=300 | 1.0000 | 1.0000 | 0.9851 | 0.9949 |
| N=500 | 1.0000 | 1.0000 | 0.9774 | 0.9982 |

Table 4. Similarities between the Estimated and True indirect Effects Matrices, with Factors

|       | Direct Effects | | Indirect Effects | |
| --- | --- | --- | --- | --- |
|       | corr2 | SSIM | corr2 | SSIM |
| N=30  | 1.0000 | 1.0000 | 0.6402 | 0.8410 |
| N=300 | 1.0000 | 1.0000 | 0.5367 | 0.9948 |
| N=500 | 1.0000 | 1.0000 | 0.8013 | 0.9982 |

It is also worthwhile pointing out that both the negative and positive spillover effects are recovered without problems. The implication of this finding is profound: the unrestricted model we have proposed is able to capture spillover effects of different signs if that is what is in the data.

### 3.3 Latent Factors

Figures 13–15 show the true factors (in blue) plotted against the estimated factors (in red). In all cases, the estimated factors closely track the true factors. This visual impression is confirmed by Table 5, which reports correlations between the true and estimated factors exceeding 0.955, suggesting that the estimates align closely with the true ones.

Table 5. Correlations between the Estimated and True Factors

|          | Factor 1 | Factor 2 | Factor 3 |
|----------|----------|----------|----------|
| N=30     | 0.9961   | 0.9901   | 0.9551   |
| N=300    | 0.9947   | 0.9669   | 0.9846   |
| N=500    | 0.9998   | 0.9771   | 0.9920   |

### 4. Application to EU regional GVA Data

Our dataset contains 202 EU NUTS2 regions listed in the Data Appendix. We use annual data spanning from 1999 to 2019, including the annual estimates of NUTS2 regional Gross Value Added (GVA), the number of scientists and engineers (sci), the number of working age population with lower education attainment (low_edu), the number of working age population with higher education attainment (high_edu), gross fixed capital (cap), total employment (emp) and total population (pop).[11]

We consider the following dynamic spatial panel model

$$y_t = W^* y_t + X_t \beta + e_t \tag{13}$$

---

[11] The datasets are from ARDECO Database and data.europa.eu - the official portal for European data.

where $y_t$ is a $202 \times 1$ vector of annual GVA growth rates, $X_t$ is a $202 \times 8$ matrix containing the lagged GVA growth rates, the initial values of GVA, and the growth rates of sci, low_edu, high_edu, capital, employment and population. All the variables are normalised for each region.

We assume that there are four latent common factors (shocks) in the error terms: the EU factor, which affects all regions; the North factor, which loads on regions in Austria, Belgium, Germany, Finland, France, Luxembourg, the Netherlands, and Sweden; the South factor, which loads on regions in Cyprus, Greece, Spain, Italy, Malta, and Portugal; and the East factor, which loads on regions in the Czech Republic, Estonia, Hungary, Latvia, Romania, Poland, and Slovakia. Furthermore, we assume that these four factors are mutually independent.

$$e_t = f_t \Lambda + \varepsilon_t \qquad (14)$$

where $\varepsilon_t \sim N(0, \Sigma)$, and $\Sigma = diag(\sigma_1^2, \dots \sigma_{202}^2)$.

We use the simple two-step VB approach to estimate the unrestricted spatial weights matrix $W^*$ and the parameter $\beta$. The coefficients of the pre-determined and exogenous variables are reported in Table 6.[12]

Table 6 Parameter Estimates

|  | Initial GVA | L GVA Growth | Scientists | Low_edu | High_edu | Capital | Emp | Population |
|---|---|---|---|---|---|---|---|---|
| Mean | -0.1527 | -0.1450 | 0.0307 | 0.0256 | -0.0164 | 0.2093 | 0.2454 | 0.0394 |
| Std | 0.0246 | 0.0242 | 0.0257 | 0.0240 | 0.0242 | 0.0236 | 0.0255 | 0.0238 |

---

[12] We have tried different lag lengths and priors and found the results to be robust.

The signs of the estimated parameters, except for those two associated with low_edu and high_edu, are in line with the findings of previous research: negative coefficients of Initial GVA and lagged GVA growth indicating GVA convergence among regions; positive coefficients of Scientist, Capital, Employment and Population in accord with economic theories. The coefficients of Low_edu and High_edu are not significant and of relatively small magnitudes; we shall refrain from over-interpreting them. But it would be interesting to investigate whether the unusual signs are caused by factors such as immigrations/emigrations and changes in high educations, in future research.

Now let us turn to the main concern of our empirical application, namely, examining the spatial interrelationships between regional GVA developments. We plot the histogram for the off-diagonal elements of the estimated spatial weights matrix in Figure 16. Clearly, a majority of the elements are of positive values and only a small number of them are negative, implying that there are more positive spillovers between regions than the negative ones.

Figure 17 plots the full estimated spatial weights matrix. It is evident that regions of the same country tend to share similar spatial interrelationship patterns. To highlight these patterns, we drop spatial elements whose absolute values are less than 0.003 from the spatial matrix and plot the remaining elements in Figure 18.

Five interesting messages emerge from Figures 17–18. First, a region tends to be more closely linked with regions within its own country than with other countries' regions. Along the zero-diagonal line, we can observe clear squared sub-matrices, associated with individual countries, whose elements are of much higher magnitudes than other elements either in the same rows or in the same columns. Second, the spatial linkages between German regions and regions of other countries are relatively loose. Only a small number of regions in Austria, Belgium, Czechia, France, Italy, Netherlands and Sweden are spatially linked with German regions. Third, in contrast to German regions which tend not to spatially interact with other countries' regions, regions in Austria,

Belgium, Spain, France, Italy and Portugal are all more closely spatially linked with each other. Fourth, regions in Greece are spatially linked with almost all regions of Spain and a small number of regions in Belgium, France, Italy and Portugal, but not much with regions of other countries. Fifth, a number of spatial elements are negative, a finding which we discuss below.

We plot the negative spatial elements in Figure 19. Five regions stand out: Pinzgau-Pongau (AT322) Salzburg und Umgebung (AT323) of Austria; Italy's Piemonte (ITC1) and Vest (RO42) of Romania. Those five regions have negative spatial relationships with almost all regions included in our sample. But the magnitudes of the spatial elements involving Germany and Sweden are much smaller than those of the elements involving other countries, suggesting that, if there are any crowding out effects, they do not affect Germany and Sweden much. If we consider smaller crowding out effect of lesser magnitudes, all regions in Germany stand out. It is seen that increases in German regions' GVA growths will negatively impact the GVA growths in Greece. Although increases in Greek regional GVA growths will negatively affect German regional GVA growths, the impact here is small in comparison with German regions' impacts on Greek regions. Apart from Greek regions, we can see regions in Germany also exert some small negative impacts on regions in Romania and Poland. It seems possible that Germany is drawing in economic activities from Greece, and to a lesser extent, Romania and Poland.

The negative relationship between German and Greek GVA growth rates is consistent with the experiences of those countries from 2001, when Greece joined the euro area, until 2015, at which time the euro-area crisis is generally considered to have ended. The period 2001 to 2015 can be separated into two parts. From 2001 to 2008, German banks greatly increased their amount of lending into the Greek banking system to take advantage of the interest-rate differential in favour of Greek financial instruments and the absence of exchange rate risk. As a result, Greek GVA growth rose relative to German GVA growth. With the outbreak of the Greek sovereign debt crisis in 2009, the direction of capital flows between the two countries reversed as Greek residents, concerned with the stability of the Greek banking system, withdrew their savings from Greek banks and deposited them into the safer German banks, reducing GVA growth in Greece and increasing GVA growth in Germany.[13]

Figures 20–22 plot the off-diagonal elements of $(I_N - W^*)^{-1}$, which govern the spillover estimates. For example, if we want to compute the spillover effects of an increase of the $i^{th}$ region's capital by 1, what we need to do is to multiply the $i^{th}$ column of $(I_N - W^*)^{-1}$ by 0.2393, the coefficient of capital, the $j^{th}$ element of the product vector will be the amount of spillover from region $i$ to region $j$. Results presented in Figures 20–22 are consistent with those presented in Figures 16–19. To save space, we do not report findings that are qualitatively the same.

Lastly, we plot the four latent common factors in Figure 23. Since all factors are normalized, meaningful comparisons can only be made for each factor over time, rather than across factors. The EU factor peaks around 2011, shortly after the global financial crisis, and reaches its lowest point in 2020, when COVID-19 struck. The North factor declines to its minimum in 2009, during the financial crisis, but remains relatively stable throughout the COVID period. The South factor peaks in 2008, just before the financial crisis took hold of Europe, and experiences the largest drop during the pandemic. The East factor shows relatively modest variation overall but increases notably after COVID.

The variance decomposition results, however, indicate that the four common factors together explain only 22.18% of the total variance in the residuals of model (13), corresponding to 7.32% of the variance in the GVA growth data. Specifically, the EU factor accounts for 7.91%, the North factor for 6.13%, the South factor for 3.86%, and the East factor for 4.29% of the residual variance. In terms of the total variance in the GVA growth rates, this translates to 2.61% explained by the EU factor, 2.02% by the North factor, 1.27% by the South factor, and 1.41% by the East factor. These results suggest that the majority of the variation is captured by the exogenous variables and spatial spillovers, rather than by the latent common shocks.

---

[13] For a detailed discussion, see Banerjee *et al.* (2021).

Our findings indicate that the spatial interdependence between EU regional economies is more complicated than those documented in the existing literature. This is potentially of substantial importance, especially to policy makers in institutions such as the European Commission and the European Investment Bank who design and implement regional policies in order to foster market integration and support economic growth.

## 5. Conclusion

If the a priori elicited spatial weights matrices are not correct, inferences of the panel SAR models can be misleading. Although there are a number of techniques to test and compare models with different weights matrices, it is usually unrealistic to assume the spatial weights matrices selected by the tests results are the 'true' ones, because, while theories and conventions can only provide abstract guidance, the number of plausible spatial weights matrices can be unlimited. Hence it is important to let the data speak for themselves.

This paper extends Ahrens and Bhattacharjee (2015) and Gefang et al. (2023). Our main purpose is to uncover the spatial interdependence between cross-sectional units without imposing any a priori spatial weights matrices. We introduce a simple two-step VB approach to estimating unrestricted panel SAR models where $N \gg T$. Using D-L prior, we are able to uncover the spatial linkage structure super-fast. Extensive Monte Carlo experiments show that our method works well. Keeping $T = 20$, we are able to identify the spatial weight matrices and spillovers that are highly correlated with their true counterparts for $N$ as large as 500, and potentially even larger N. Furthermore, allowing for more complicated unobserved heterogeneity in the error terms, we show that available Bayesian techniques can be used to extract the latent factors from the residuals.

In empirical applications, we investigate the spatial interdependencies of EU regional GVA growth. Apart from identifying prominent country clusters, we are also able to reveal a number of spatial relationships that have not been uncovered before, providing important information to support policy makers' decision.

## Appendix. List of NUTS2 regions

| Austria | Germany | Greece | France | Hungary | Poland | Slovakia |
|---|---|---|---|---|---|---|
| AT11 | DE11 | EL30 | FR10 | HU21 | PL21 | SK01 |
| AT12 | DE12 | EL41 | FRB0 | HU22 | PL22 | SK02 |
| AT13 | DE13 | EL42 | FRC1 | HU23 | PL41 | SK03 |
| AT21 | DE14 | EL43 | FRC2 | HU31 | PL42 | SK04 |
| AT22 | DE21 | EL51 | FRD1 | HU32 | PL43 | |
| AT31 | DE22 | EL52 | FRD2 | HU33 | PL51 | |
| AT32 | DE23 | EL53 | FRE1 | **Italy** | PL52 | |
| AT33 | DE24 | EL54 | FRE2 | ITC1 | PL61 | |
| AT34 | DE25 | EL61 | FRF1 | ITC2 | PL62 | |
| **Belgium** | DE26 | EL62 | FRF2 | ITC3 | PL63 | |
| BE10 | DE27 | EL63 | FRF3 | ITC4 | PL71 | |
| BE21 | DE30 | EL64 | FRG0 | ITF1 | PL72 | |
| BE22 | DE40 | EL65 | FRH0 | ITF2 | PL81 | |
| BE23 | DE50 | **Spain** | FRI1 | ITF3 | PL82 | |
| BE24 | DE60 | ES11 | FRI2 | ITF4 | PL84 | |
| BE25 | DE71 | ES12 | FRI3 | ITF5 | **Malta** | |
| BE31 | DE72 | ES13 | FRJ1 | ITF6 | MT00 | |
| BE32 | DE73 | ES21 | FRJ2 | ITG1 | **Portugal** | |
| BE33 | DE80 | ES22 | FRK1 | ITG2 | PT11 | |

| | | | | | |
|---|---|---|---|---|---|
| BE34 | DE91 | ES23 | FRK2 | ITH1 | PT15 |
| BE35 | DE92 | ES24 | FRL0 | ITH2 | PT16 |
| Cyprus | DE93 | ES30 | Netherlands | ITH3 | PT17 |
| CY00 | DE94 | ES41 | NL11 | ITH4 | PT18 |
| Czechia | DEA1 | ES42 | NL12 | ITI1 | PT20 |
| CZ01 | DEA2 | ES43 | NL13 | ITI2 | Sweden |
| CZ02 | DEA3 | ES51 | NL21 | ITI4 | SE11 |
| CZ03 | DEA4 | ES52 | NL22 | Romania | SE12 |
| CZ04 | DEA5 | ES53 | NL23 | RO11 | SE21 |
| CZ05 | DEB1 | ES61 | NL31 | RO12 | SE22 |
| CZ06 | DEB2 | ES62 | NL32 | RO21 | SE23 |
| CZ07 | DEB3 | ES63 | NL33 | RO22 | SE31 |
| CZ08 | DEC0 | ES64 | NL34 | RO31 | SE32 |
| Estonia | DED2 | ES70 | NL41 | RO32 | SE33 |
| EE00 | DEE0 | Finland | NL42 | RO41 | |
| Luxembourg | DEF0 | FI19 | Latvia | RO42 | |
| LU00 | DEG0 | FI20 | LV00 | | |

Footnote: The NUTS classifications can be found at https://ec.europa.eu/eurostat/web/nuts/background

Figure 1. Spatial Weights Matrices for $N = 30$, without Factors

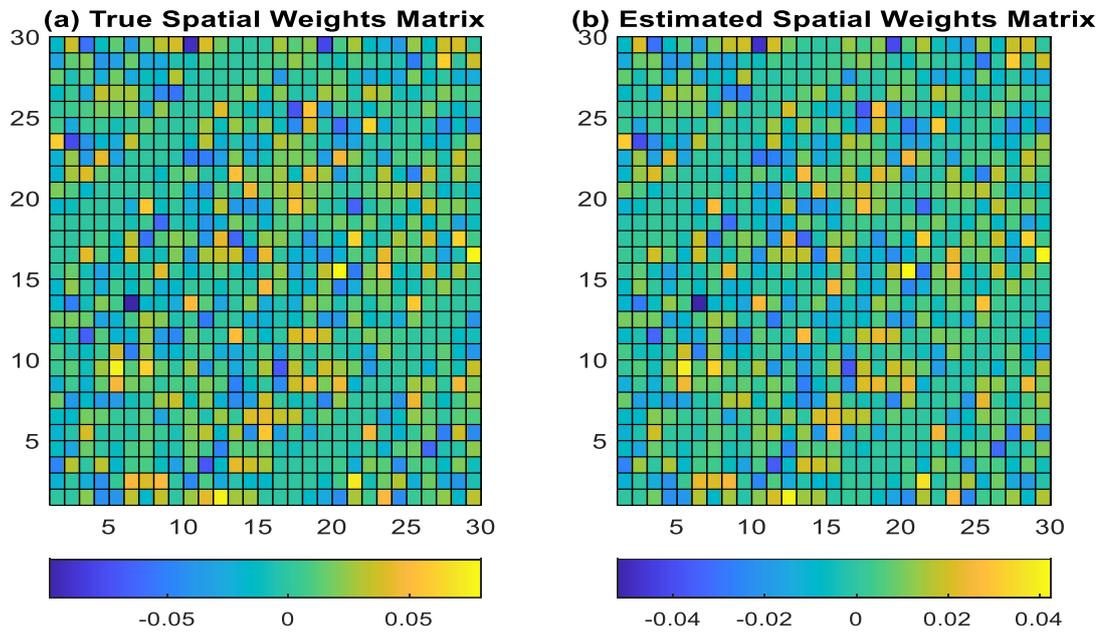

Figure 2. Spatial Weights Matrices for $N = 30$, with Factors

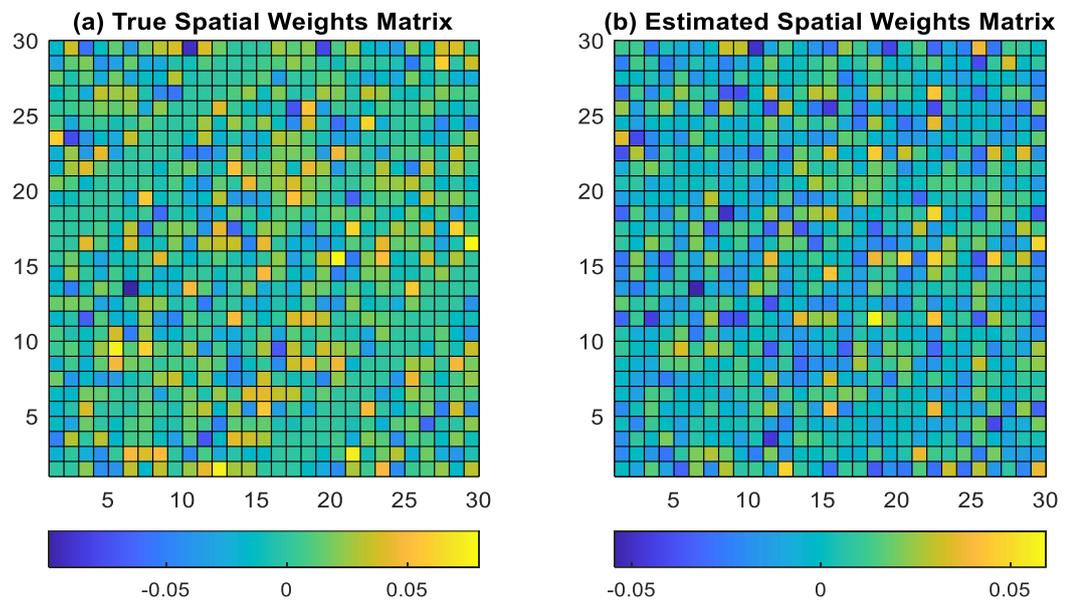

Figure 3. Spatial Weights Matrix for $N = 300$, without Factors

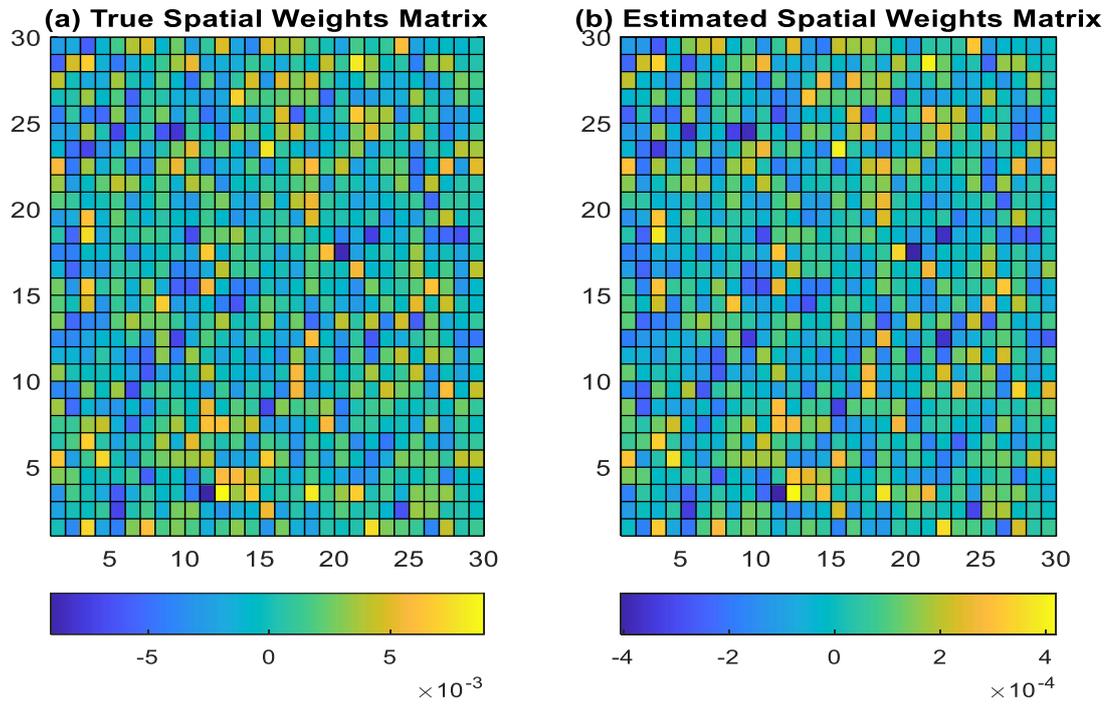

Figure 4. Spatial Weights Matrix for $N = 300$, with Factors

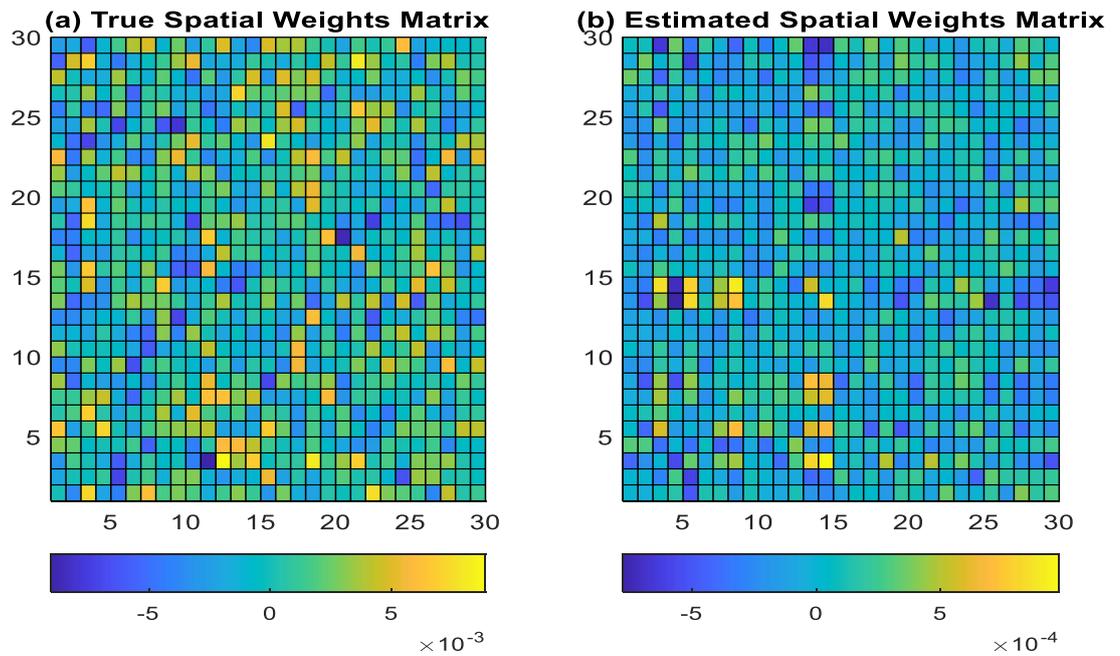

Figure 5. Spatial Weights Matrices for $N = 500$, without Factors

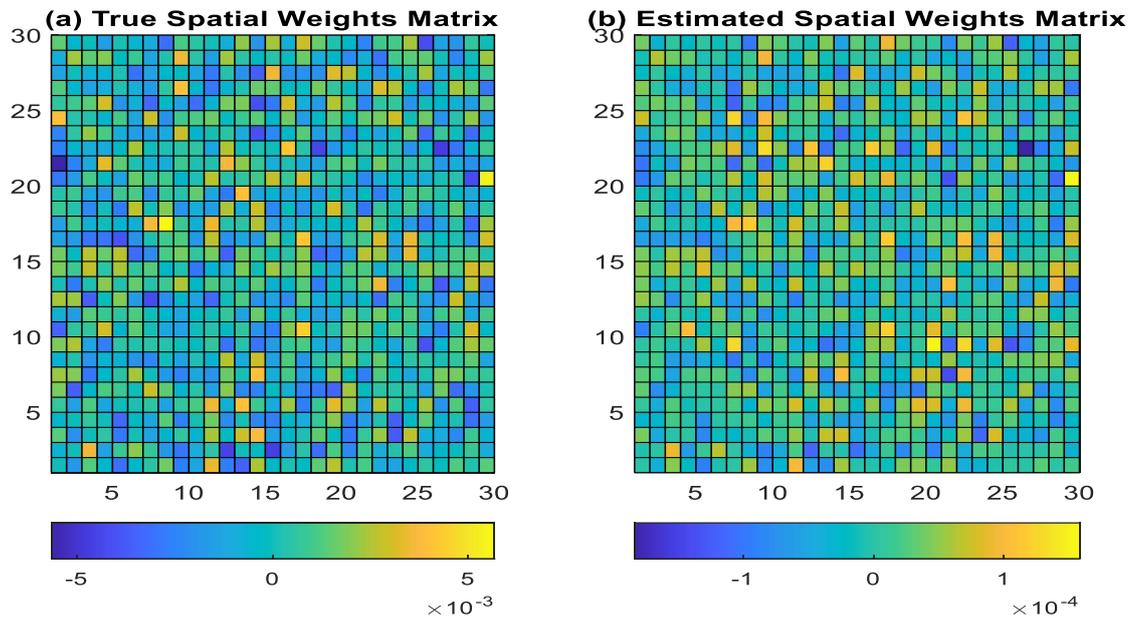

Figure 6. Spatial Weights Matrices for $N = 500$, with Factors

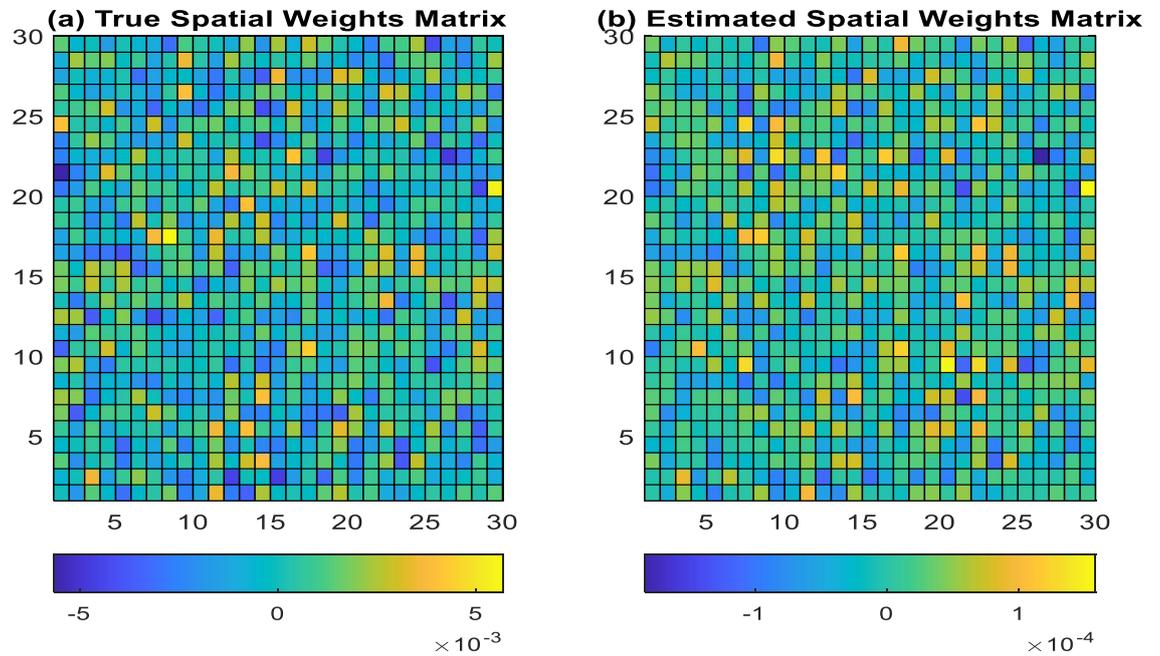

Figure 7. Effects Matrices for $N = 30$, without Factors

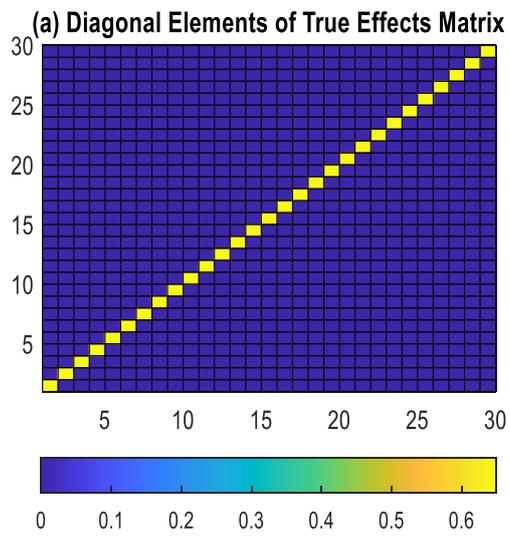

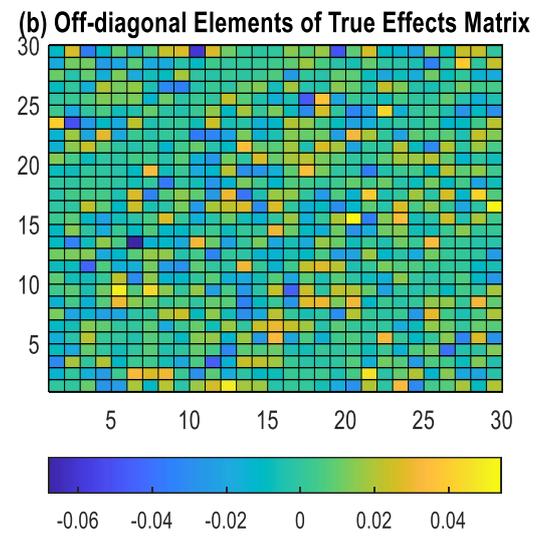

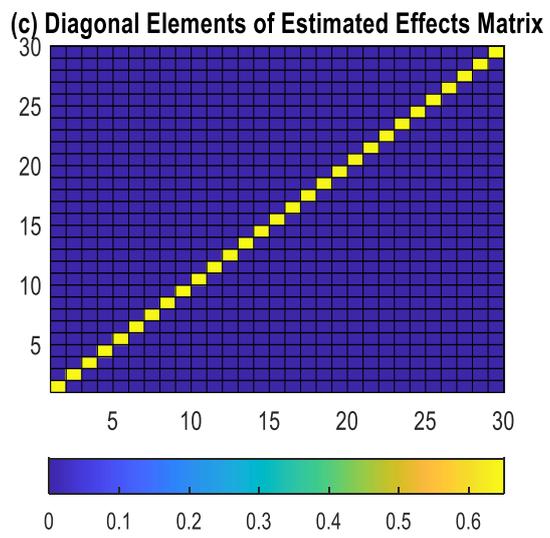

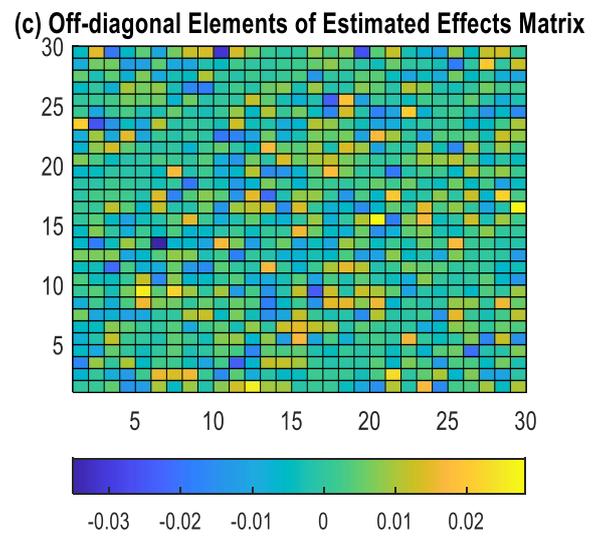

Figure 8. Effects Matrices for $N = 30$, with Factors

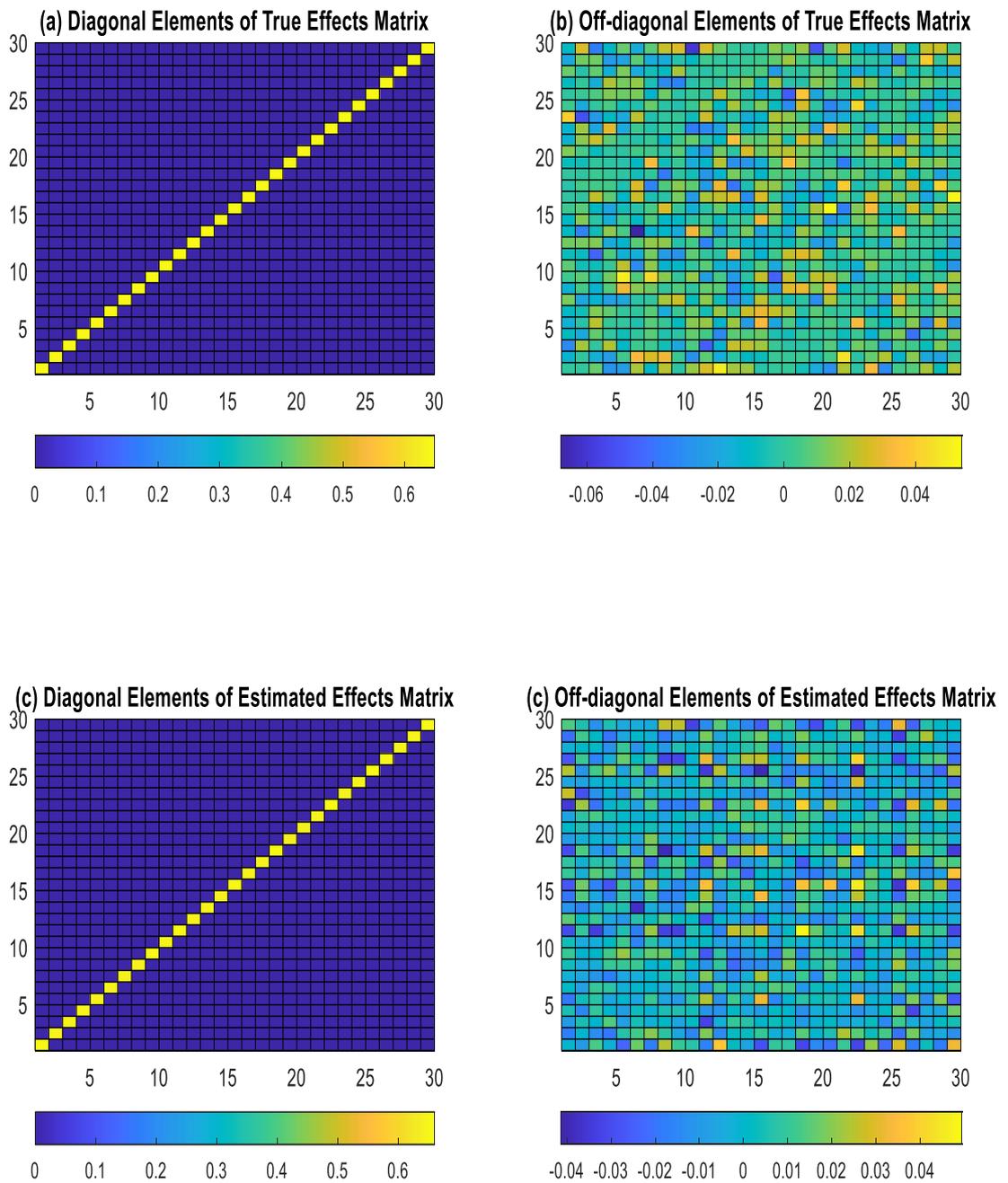

Figure 9. Effects Matrices for $N = 300$, without Factors

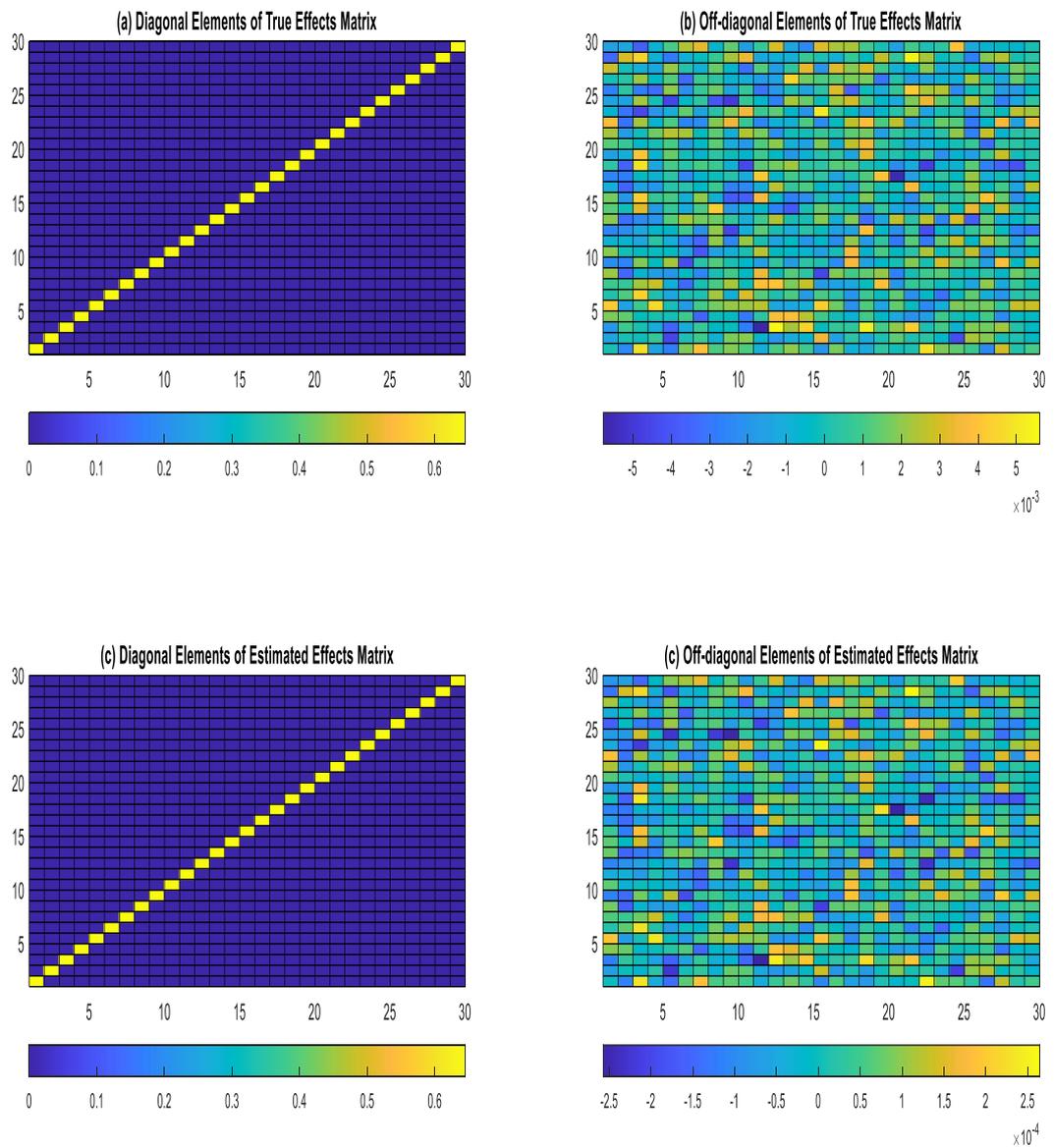

Figure 10. Effects Matrices for $N = 300$, with Factors

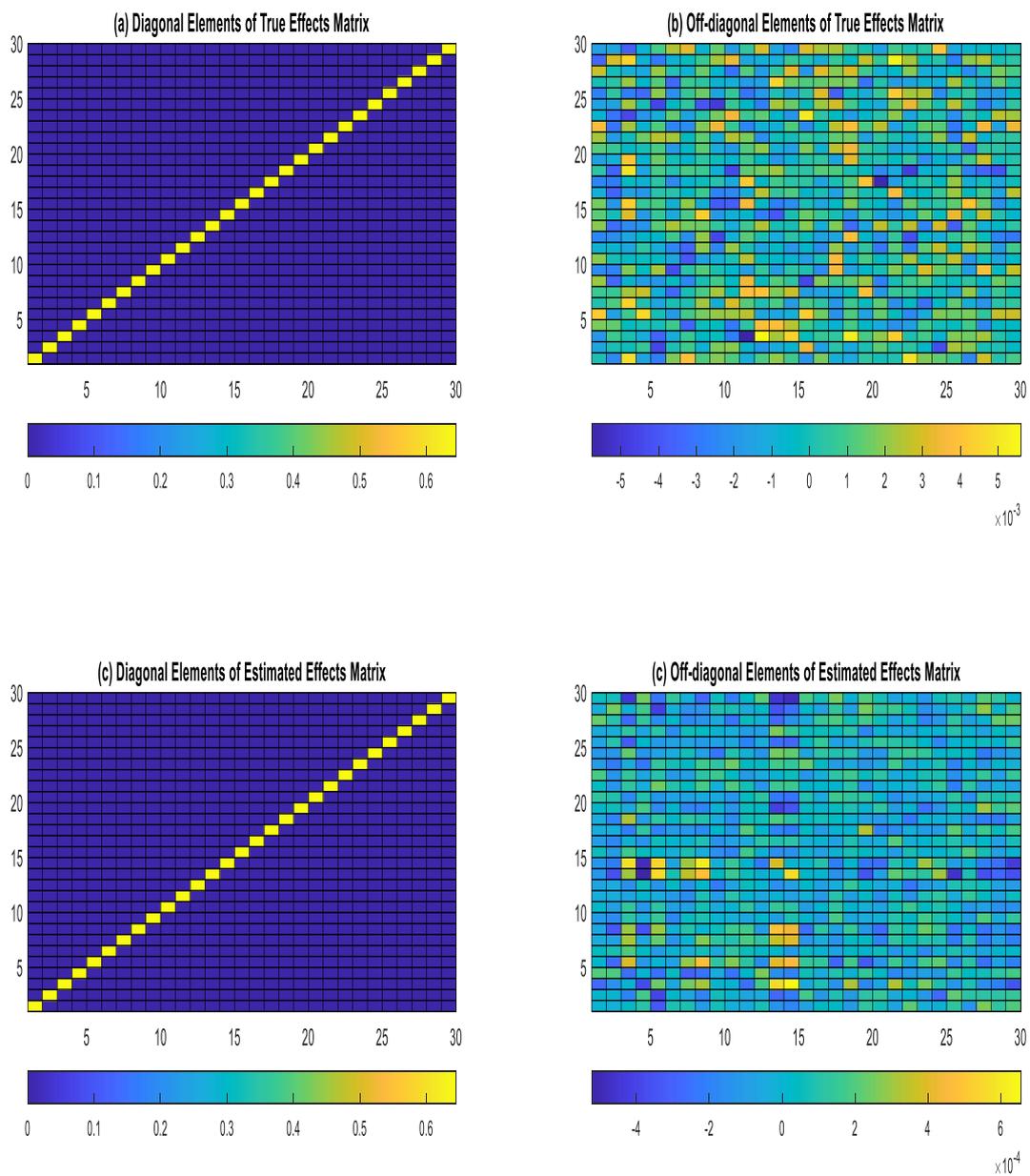

Figure 11. Effects Matrices for $N = 500$, without Factors

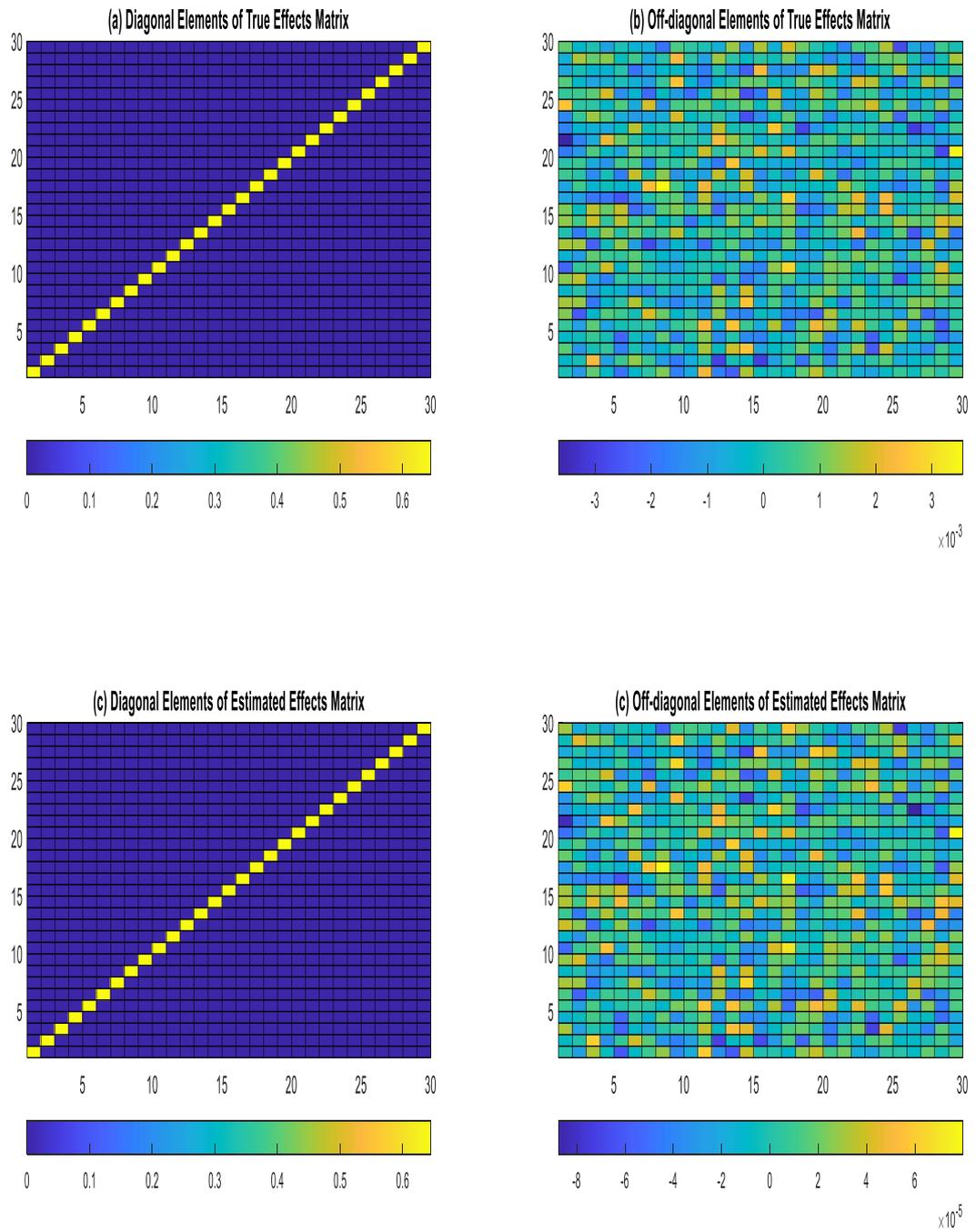

Figure 12. Effects Matrices for $N = 500$, with Factors

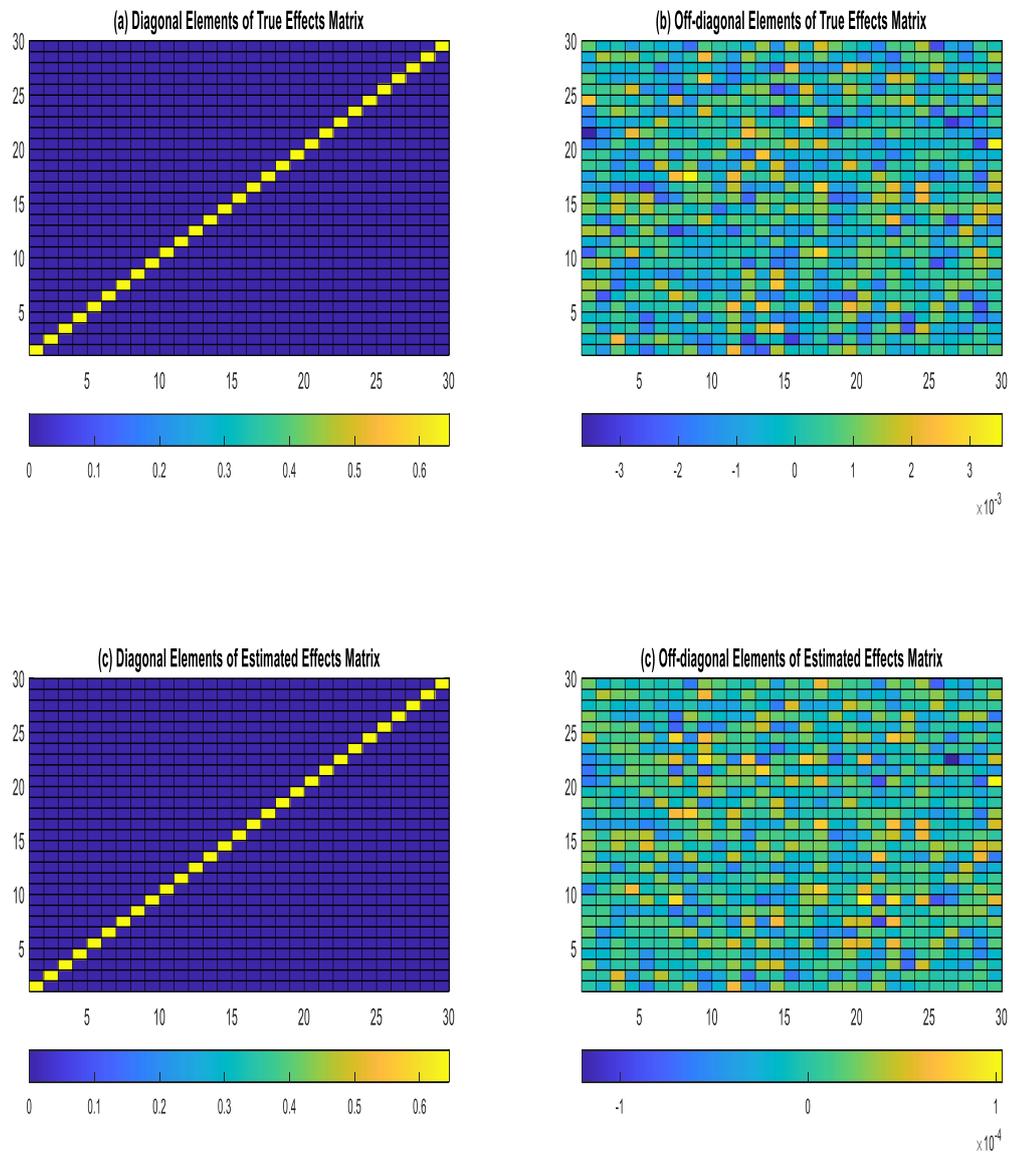

Figure 13. Factors for $N = 30$

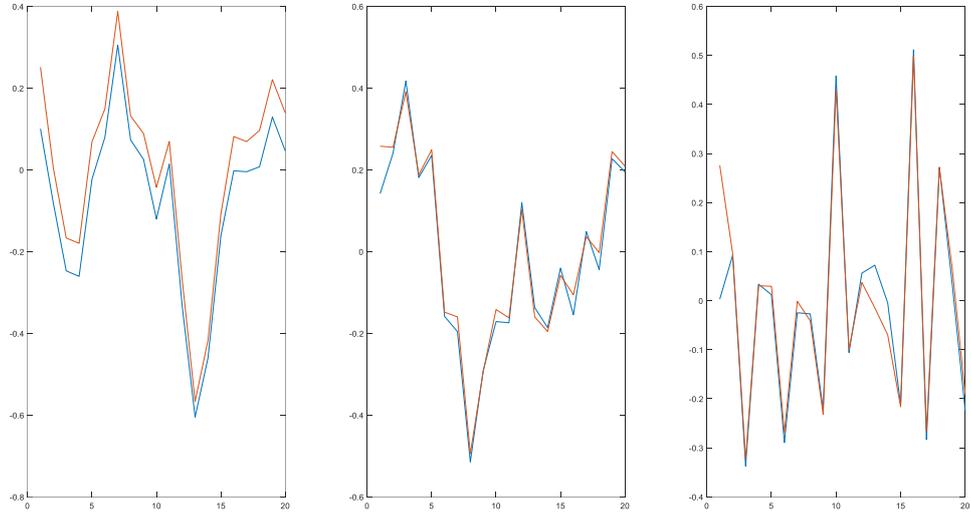

Figure 14. Factors for $N = 300$

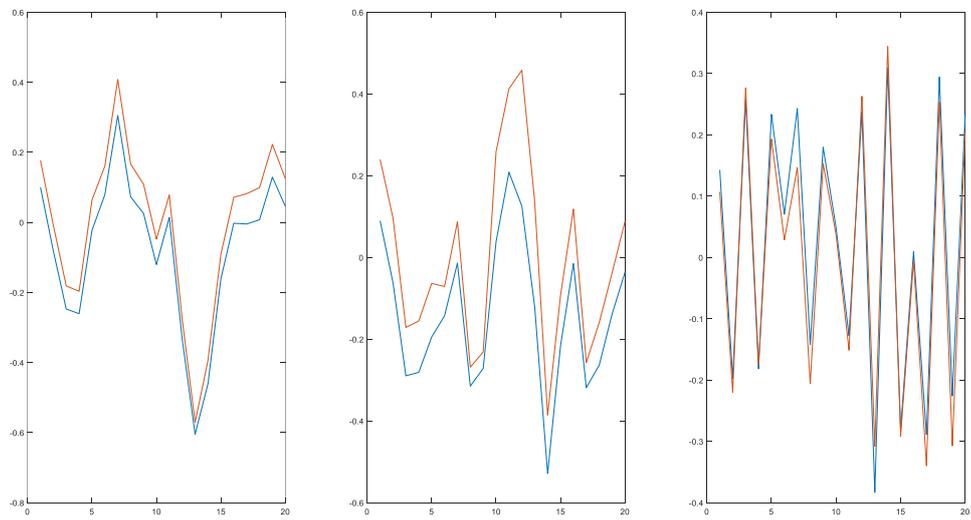

Figure 15. Factors for $N = 500$

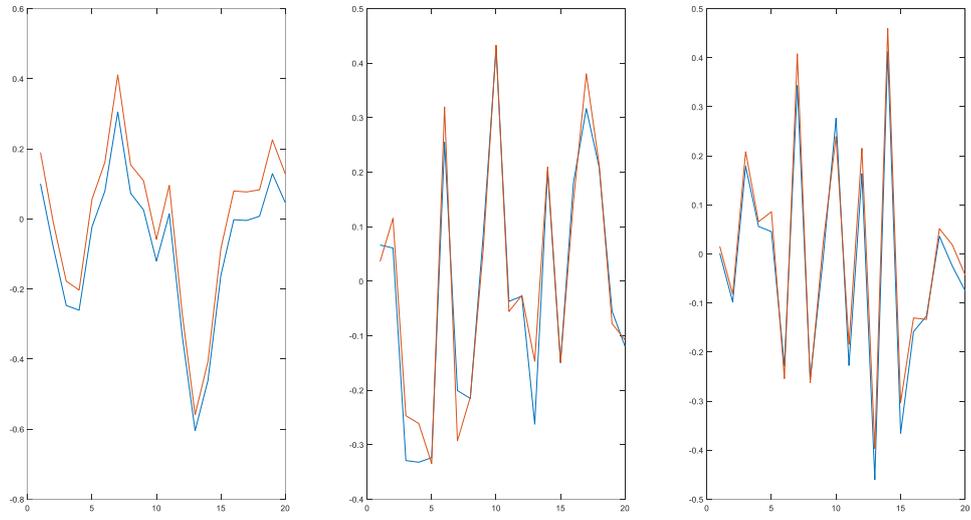

Figure 16. Histogram of the off-diagonal elements of the estimated spatial weights matrix for GVA data

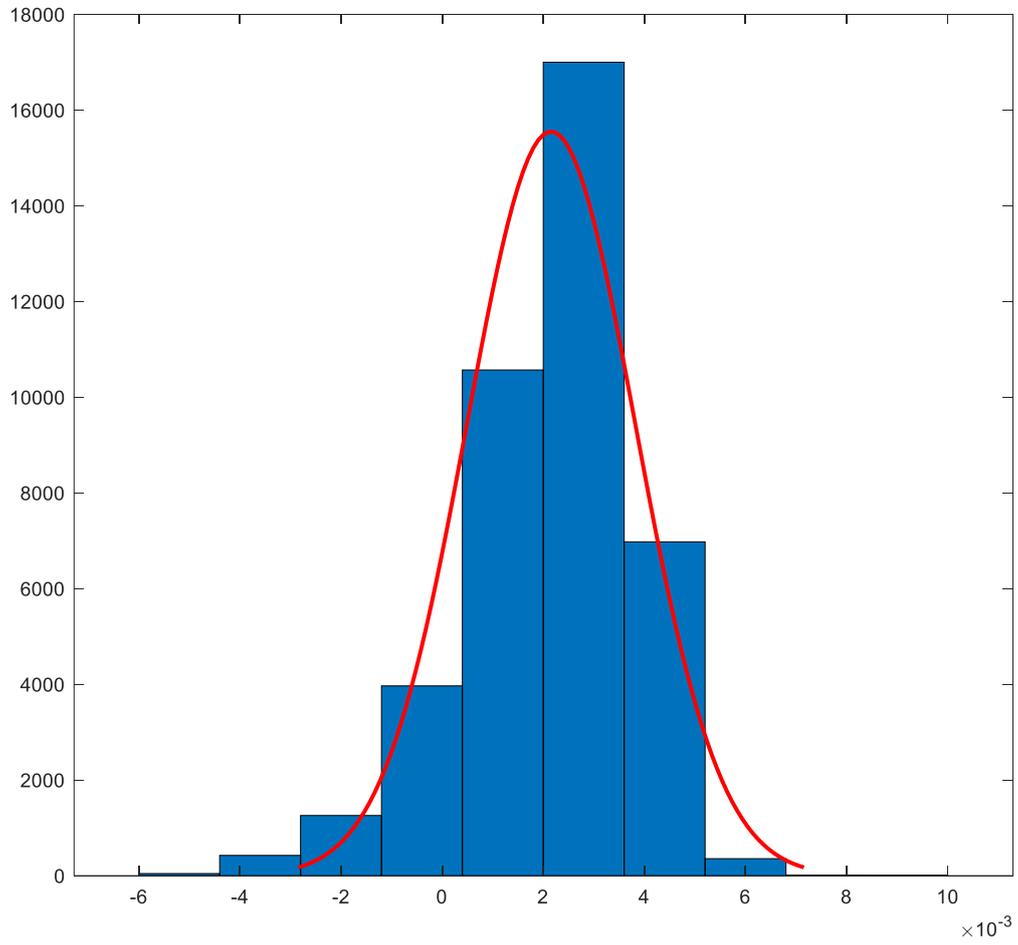

Figure 17. Estimated Spatial Weights Matrix for GVA Data (i),

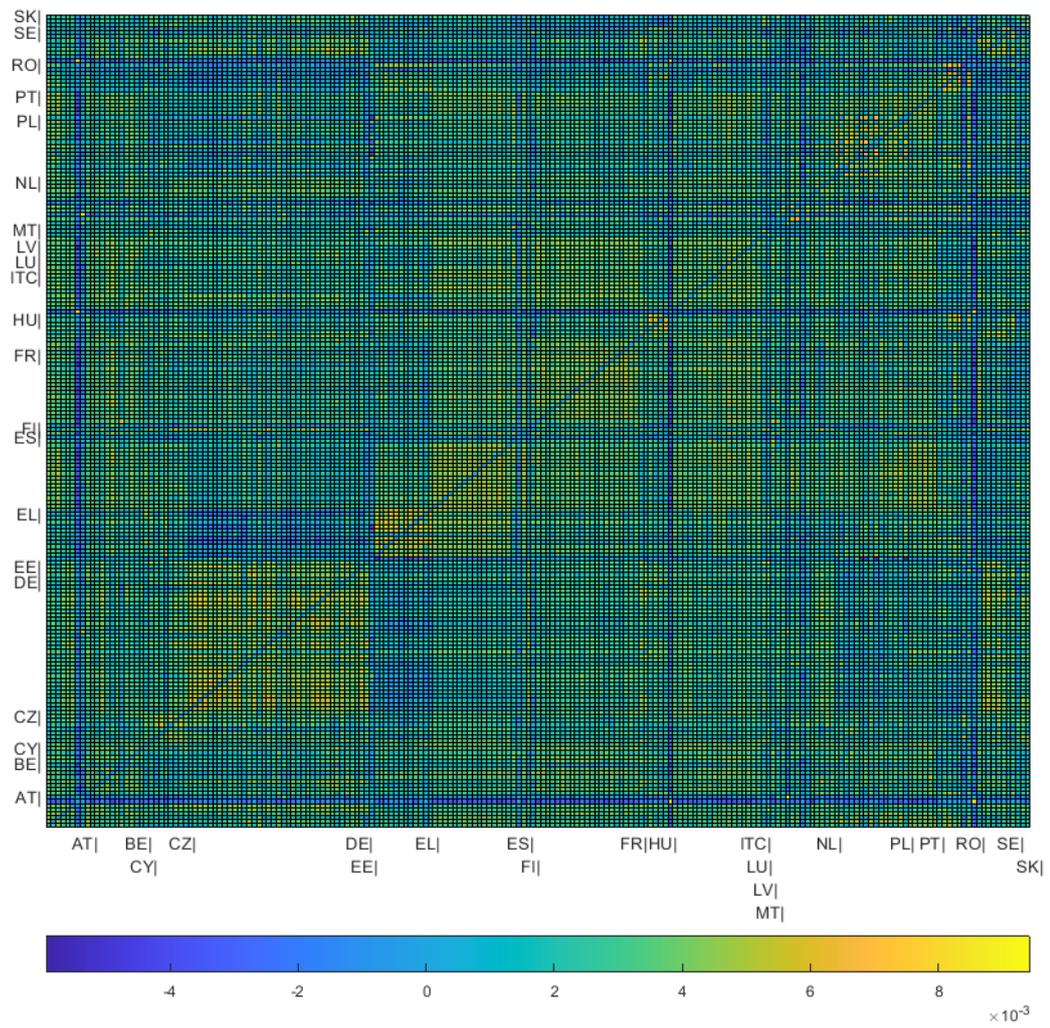

Figure 18. Estimated Spatial Weights Matrix for GVA Data (ii)

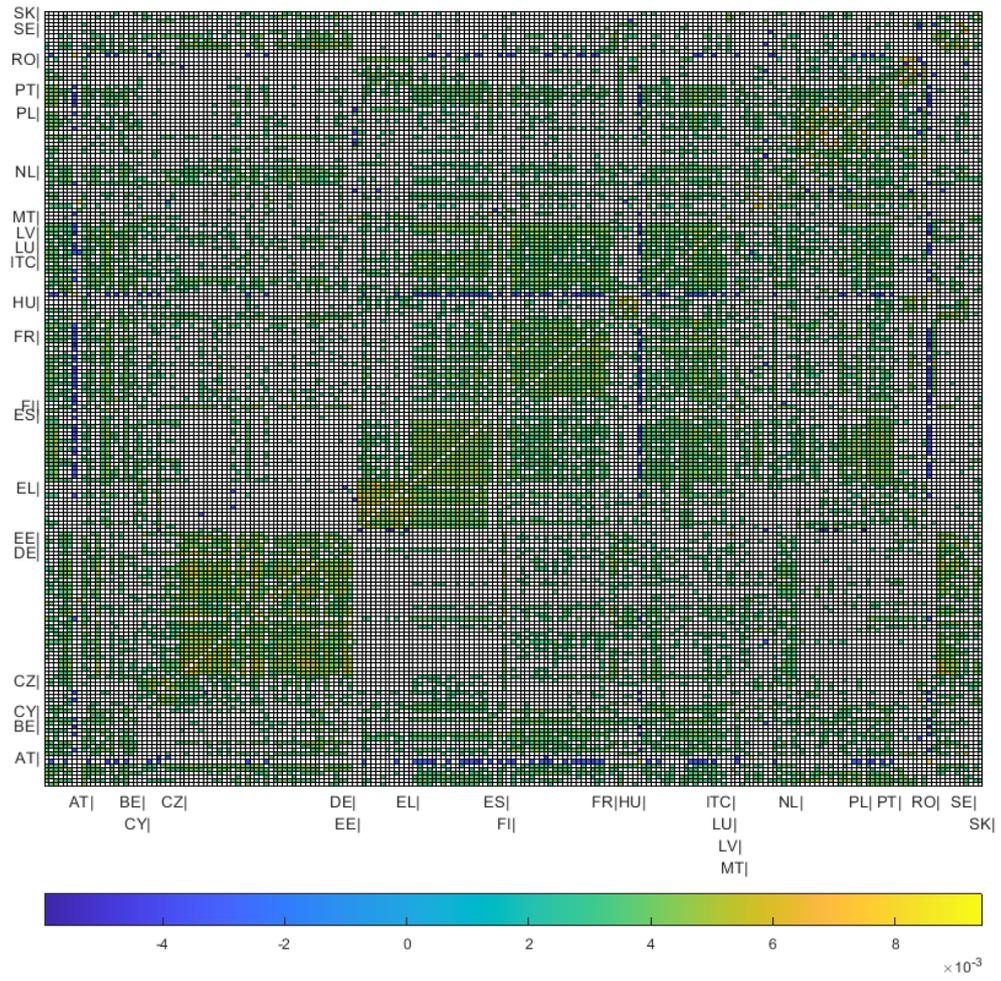

Figure 19. Estimated Spatial Weights Matrix for GVA Data (iii)

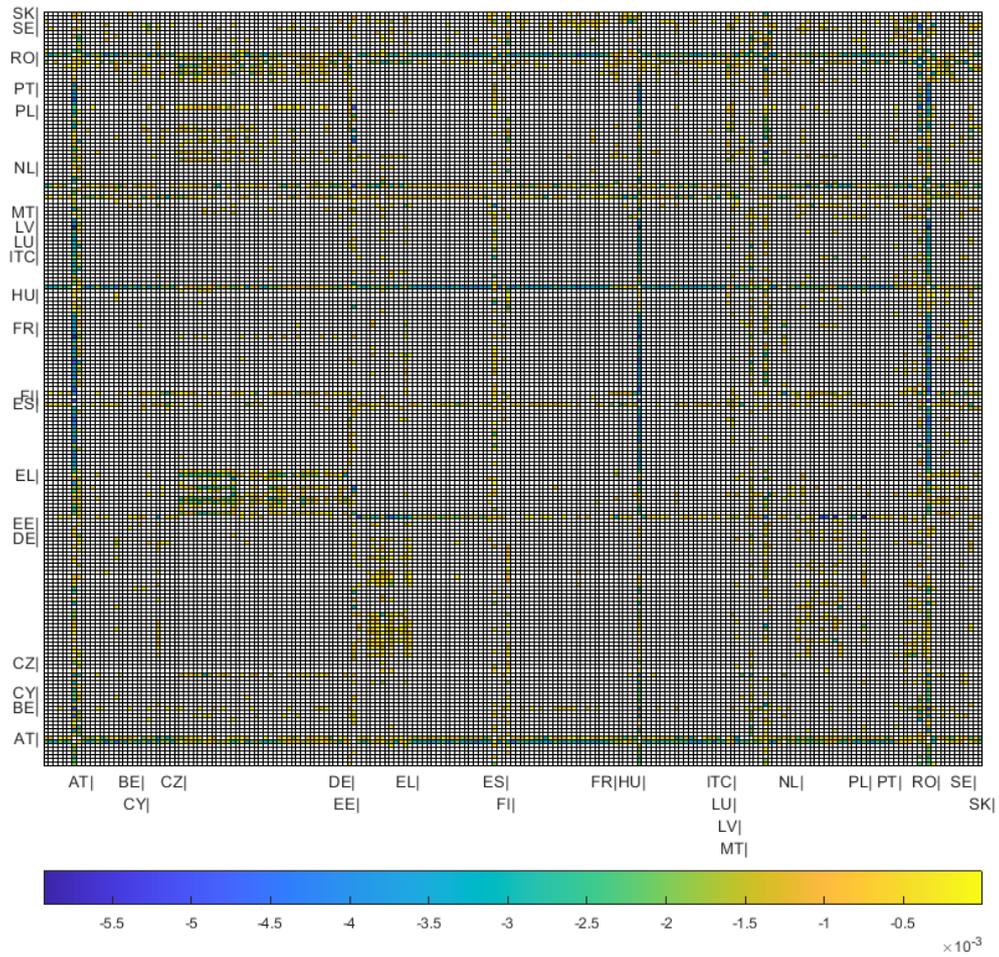

Figure 20. Histogram of the Off-diagonal Elements in $(I_N - W^*)^{-1}$

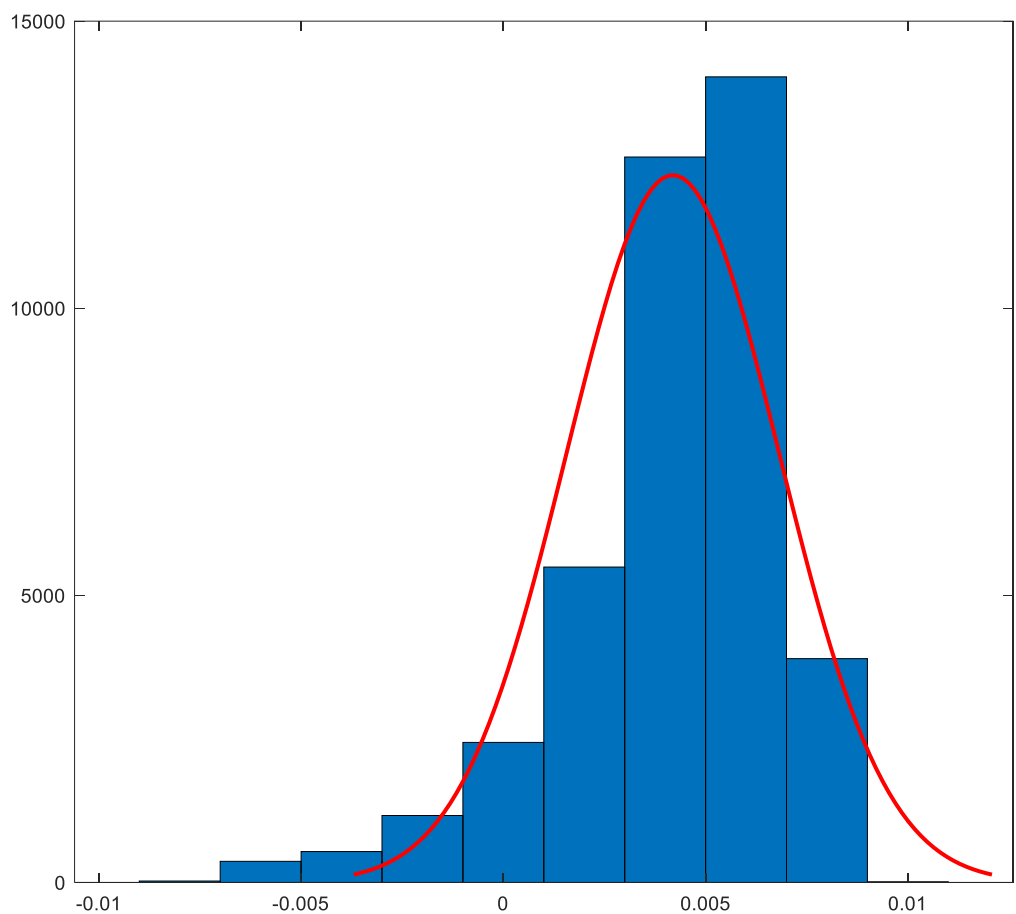

Figure 21. Matrix $(I_N - W^*)^{-1}$ with Zero Diagonal Elements (i)

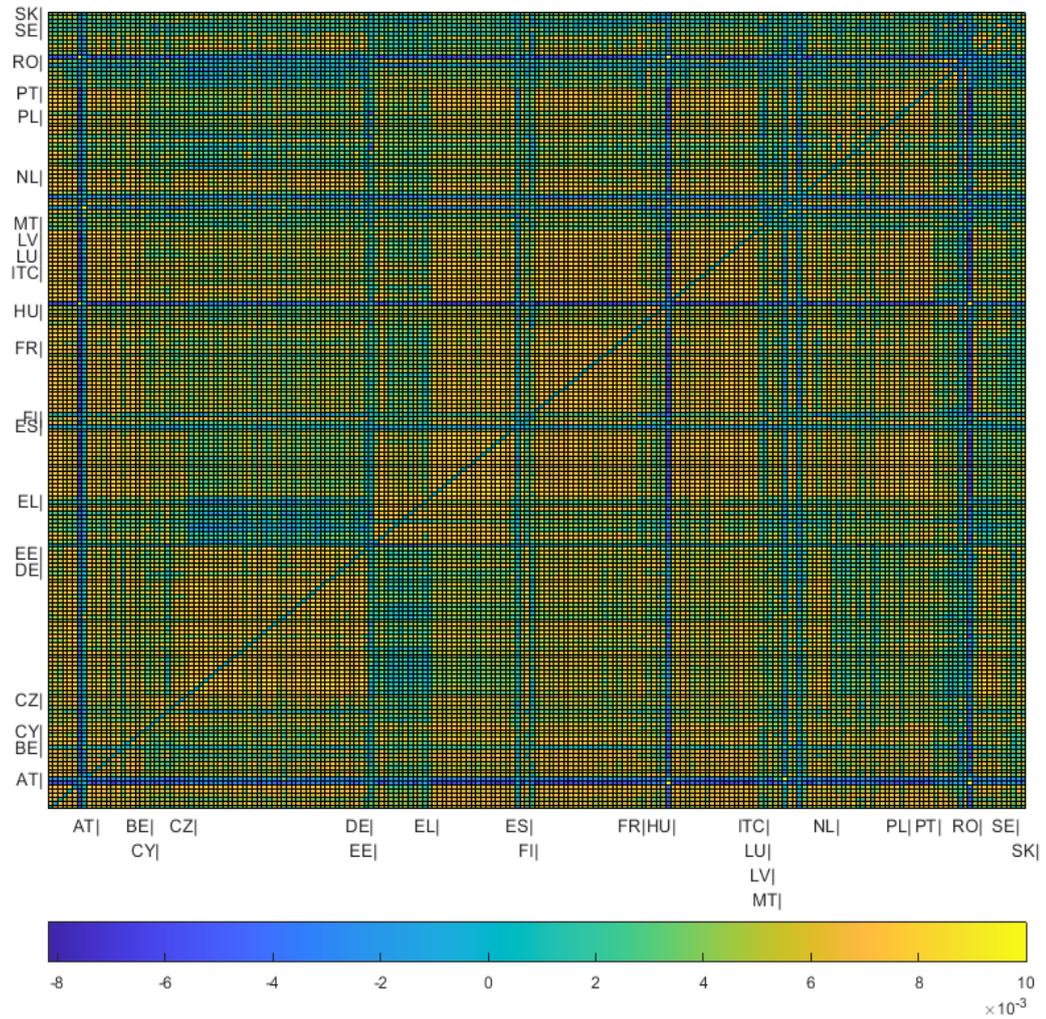

Figure 22. Matrix $(I_N - W^*)^{-1}$ with Zero Diagonal Elements (ii)

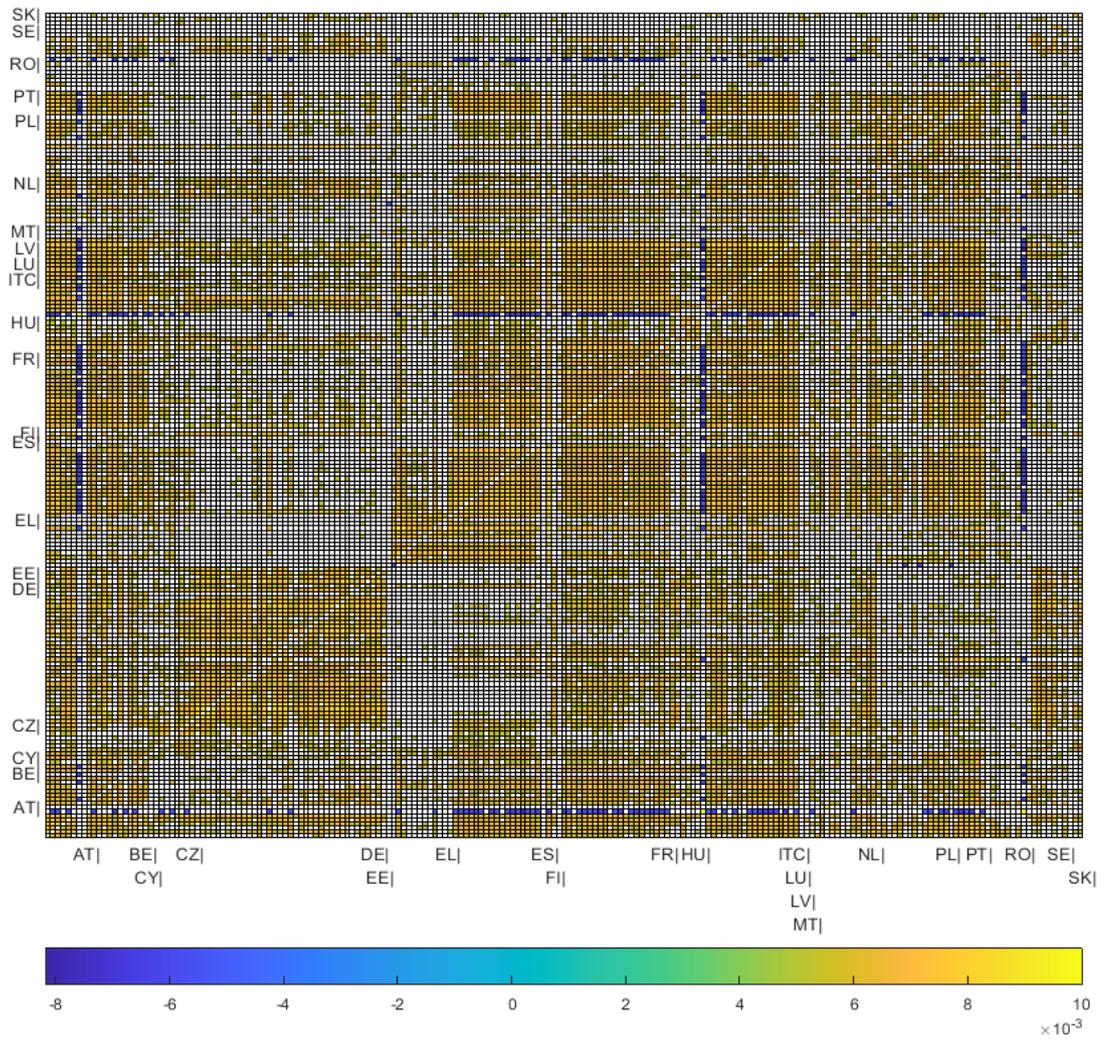

Figure 23.  Common Factors

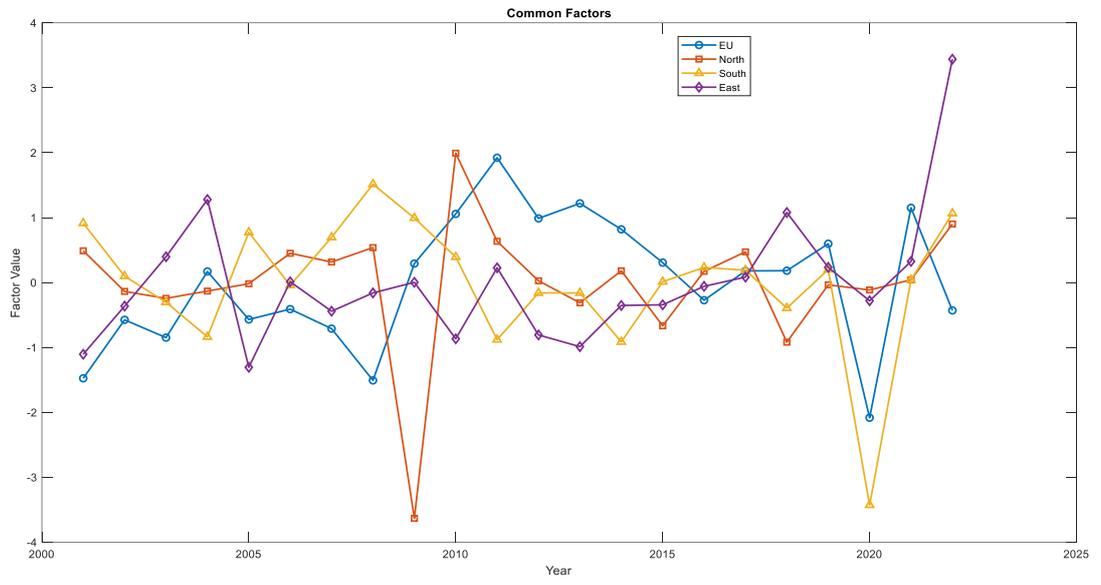